\documentclass[10pt]{article}
\usepackage{amssymb}
\usepackage[dvips]{epsfig}
\usepackage{graphicx}

\newtheorem{Lemma}{Lemma}[section]

\newtheorem{Proposition}[Lemma]{Proposition}
\newtheorem{Corollary}[Lemma]{Corollary}

\newenvironment{Proof}%
 {\begin{trivlist} \item[]{\bf Proof. }}%
 {\hspace*{\fill}$\rule{.3\baselineskip}{.35\baselineskip}$\end{trivlist}}
 {\begin{trivlist} \item[]{\bf Proof}}%
 {\hspace*{\fill}$\rule{.3\baselineskip}{.35\baselineskip}$\end{trivlist}}

\renewcommand{\theLemma}{\arabic{section}.\arabic{Lemma}}

\makeatletter \@addtoreset{equation}{section} \makeatother

\newcommand{\C}{\mathbb{C}}

\newcommand{\R}{\mathbb{R}}

\def\Re{\mathop{\mathrm{Re}}}

\newfam\bifam
\font\tenbi=cmmib10 scaled \magstep1 \font\sevenbi=cmmib10 at 11pt
\font\fivebi=cmmib10 at 6pt \textfont\bifam = \tenbi
\scriptfont\bifam= \sevenbi \scriptscriptfont\bifam= \fivebi

\begin{document}

\title{\bf Instabilities of multi-hump vector solitons  \\
in coupled nonlinear Schr\"{o}dinger equations }

\author{Dmitry E. Pelinovsky$^\dagger$, \hspace{0.2cm}
Jianke Yang$^{\dagger\dagger}$\\  {\small $^\dagger$ Department of
Mathematics, McMaster University, Hamilton, Ontario, Canada, L8S
4K1}\\  {\small $^{\dagger\dagger}$ Department of Mathematics,
University of Vermont, Burlington, VT 05401, USA} }

\date{}
\maketitle

\begin{abstract}
Spectral stability of multi-hump vector solitons in the Hamiltonian
system of coupled nonlinear Schr\"{o}dinger (NLS) equations is
investigated both analytically and numerically. Using the closure
theorem for the negative index of the linearized Hamiltonian, we
classify all possible bifurcations of unstable eigenvalues in the
systems of coupled NLS equations with cubic and saturable
nonlinearities. We also determine the eigenvalue spectrum
numerically by the shooting method. In case of cubic nonlinearities,
all multi-hump vector solitons in the non-integrable model are found
to be linearly unstable. In case of saturable nonlinearities, stable
multi-hump vector solitons are found in certain parameter regions,
and some errors in the literature are corrected.
\end{abstract}

\section{Introduction}

The coupled NLS equations have wide applications in the modeling of
physical processes. For instance, such equations with the cubic
nonlinearity govern the nonlinear interaction of two wave packets
\cite{Benney} and optical pulse propagation in birefringent fibers
\cite{Menyuk} or wavelength-division-multiplexed optical systems
\cite{Agrawal,Hasegawa}. Similar equations with the saturable
nonlinearity describe the propagation of several mutually-incoherent
laser beams in biased photorefractive crystals
\cite{Chris_Segev96,Kivshar_book}. Various types of vector solitons
including single-hump and multi-hump ones have been known to exist
in these coupled NLS equations
\cite{Akhmediev95,Akhmediev,Champ,Christodoulides88,Chris_Segev96,Haelterman,Kivshar_book,KivOstrov,Snyder94,Tratnik,Yang},
and they have been observed in photorefractive crystals as well
\cite{ChenOL96,ChenOL00,Mitchell98}.

Linear stability of vector solitons in the coupled NLS equations is
an important issue. Fundamental single-hump vector solitons are
known to be stable \cite{KivSkryabin,PelYang,YangStudies}. Stability
of multi-hump vector solitons (which have one or more nodal points
in one or more components) is more subtle. For the cubic
nonlinearity, it was conjectured in \cite{Yang} based on the
numerical evidence that multi-hump vector solitons were all linearly
unstable. If the multi-hump solitons are pieced together by a few
fundamental vector solitons, then their linear instability has been
proven both analytically and numerically in
\cite{YangPRE01,YangAMS}. The linear instability for other types of
multi-hump vector solitons has not been proven yet. For the
saturable nonlinearity, multi-hump solitons have been shown to be
stable in certain parameter regions \cite{KivOstrov,KivSkryabin},
but the origins of their stability and instability have not yet been
fully analyzed.

>From a broader point of view, the theory of linear stability of
vector solitons in coupled NLS equations was recently developed with
the use of the closure theorem for the negative index of the
linearized Hamiltonian \cite{KKS04,P03}. However, there are not many
applications of the general theory to particular bifurcations of
unstable eigenvalues \cite{KK04,PY03}, because the general theory
excludes non-generic bifurcations. It is desirable to further
develop a perturbation theory to the eigenvalue bifurcations, so
that the origin of instability becomes more apparent in the context
of the closure theorem.

In this paper, we investigate the linear stability of multi-hump
vector solitons in the general Hamiltonian system of coupled NLS
equations both analytically and numerically. Using the closure
theorem for the negative index of the linearized Hamiltonian as well
as the perturbation technique, we classify all possible bifurcations
of unstable eigenvalues in two physical models with cubic or
saturable nonlinearities. In the first model, we show that
multi-hump vector solitons near points of local bifurcations are
always linearly unstable, in agreement with numerical results in
\cite{Yang}. In the second model, the situation is more complicated.
Our results show that the 1st family of multi-hump vector solitons
is indeed linearly stable near the local-bifurcation boundary, in
agreement with numerical results in \cite{KivSkryabin}. However, for
the 2nd family, we discovered a new oscillatory instability near the
local-bifurcation boundary, which was missed in \cite{KivSkryabin}.
Due to this oscillatory instability, the stability region of vector
solitons for the 2nd family is drastically reduced from that
reported in \cite{KivSkryabin}. Numerically, we track the unstable
eigenvalues of multi-hump solitons and reveal various scenarios of
eigenvalue bifurcations away from the local-bifurcation boundaries.
We also map out the correct stability regions of multi-hump vector
solitons in the entire parameter space. Furthermore, the number of
numerically-obtained unstable eigenvalues agrees completely with
that predicted by the negative index of the linearized Hamiltonian.

Our paper is structured as follows. The main formalism and the
closure theorem for the negative index of the linearized Hamiltonian
are described in Section 2. Analysis of unstable eigenvalues in the
coupled NLS equations with cubic and saturable nonlinearities is
developed in Sections 3 and 4, respectively. Section 5 summarizes
our results and open questions. Appendix A reviews bifurcations of
unstable eigenvalues by the perturbation method.

\section{Main formalism}

We consider a general Hamiltonian system of coupled NLS equations in
the form:
\begin{equation}
\label{NLS} i \frac{\partial \psi_n}{\partial z} + d_n
\frac{\partial^2 \psi_{n}}{\partial x^2} + \frac{\partial
U}{\partial |\psi_n|^2} \psi_n = 0, \;\;\;\; n = 1,...,N,
\end{equation}
where $z \in \R_+$, $x \in \R$, $\psi_n \in \C$, $d_n \in \R$, and
$U = U(|\psi_1|^2,...,|\psi_N|^2) \in \R$. We assume that $U(0) =
U'(0) = 0$, and $d_n > 0$ for all $n$. In optical fibers
(photorefractive crystals), the function $\psi_n(z,x)$ is the
envelope amplitude of the $n^{\rm th}$ channel (beam), $z$ is the
propagation distance along the fiber (waveguide), and $x$ is the
retarded time (the transverse coordinate)
\cite{Chris_Segev96,Hasegawa,Kivshar_book,Menyuk}.

Following the recent work in \cite{P03}, we study the linear
stability of vector solitons:
\begin{equation}
\label{soliton} \psi_n(z,x) = \Phi_n(x)  e^{i \beta_n z},
\end{equation}
where $\Phi_n : \R \rightarrow \R$, and $\beta_n > 0$ for all $n$.
We assume that none of the components $\Phi_n(x)$ vanish identically
on $x \in \R$. Linearization of the coupled NLS equations
(\ref{NLS}) follows from the expansion:
\begin{equation}
\label{expansion} \psi_n(z,x) = \left\{ \Phi_n(x) + \left[ u_n(x) +
i w_n(x) \right] e^{\lambda z} + \left[ \bar{u}_n(x) + i
\bar{w}_n(x) \right] e^{\bar{\lambda} z} \right\} e^{i \beta_n z},
\end{equation}
where $\|u_n\|, \|w_n\| \ll 1$, and the overline denotes the complex
conjugation. The linearized equations for $(u_n, w_n)$ are the
following non-self-adjoint problem in
$L^2(\mathbb{R},\mathbb{C}^{2N})$:
\begin{equation}
\label{eigenvalue-problem} \label{L} {\cal L}_1 {\bf u} = -
\lambda {\bf w}, \qquad {\cal L}_0 {\bf w} = \lambda {\bf u},
\end{equation}
where $\lambda \in \C$ is an eigenvalue, $({\bf u},{\bf w})^T : \R
\rightarrow \C^{2N}$ is the eigenvector, and ${\cal L}_0$ and ${\cal
L}_1$ are the matrix Schr\"{o}dinger operators with elements:
\begin{eqnarray}
\label{L0} ({\cal L}_0)_{n,m} & = & \left( - d_n \frac{d^2}{d x^2}
+ \beta_n - \frac{\partial U}{\partial \Phi_n^2} \right)
\delta_{n,m}, \\
 \label{L1} ({\cal L}_1)_{n,m} & = &\left( - d_n \frac{d^2}{d x^2}
+ \beta_n - \frac{\partial U}{\partial \Phi_n^2} \right)
\delta_{n,m} - 2 \frac{\partial^2 U}{\partial \Phi_n^2 \partial
\Phi_m^2} \Phi_n  \Phi_m.
\end{eqnarray}

Since the eigenvalue problem (\ref{eigenvalue-problem}) is a
linearization of the Hamiltonian system, the values of $\lambda$
occur as pairs of real or purely imaginary eigenvalues, or as
quadruplets of complex eigenvalues. Eigenvalues with ${\rm
Re}(\lambda) > 0$ lead to spectral instability of vector solitons
(\ref{soliton}). We denote the number of eigenvalues in the first
open quadrant as $N_{\rm comp}$, the number of positive real
eigenvalues as $N_{\rm real}$, and the number of purely imaginary
eigenvalues with positive ${\rm Im}(\lambda)$ as $N_{\rm imag}$. The
continuous spectrum has $N$ branches, located at the positive
imaginary axis for ${\rm Im}(\lambda) \geq \beta_n$, $n = 1,...,N$.
Zero eigenvalue $\lambda = 0$ has geometric multiplicity of at least
$(N + 1)$ and algebraic multiplicity of at least $(2N + 2)$, in the
assumption that none of the components $\Phi_n(x)$ vanishes
identically on $x \in \R$ \cite{P03}.

Furthermore, we denote the number of negative and zero eigenvalues
of operators ${\cal L}_{0,1}$ in $L^2(\R,\C^N)$ as $n({\cal
L}_{0,1})$ and $z({\cal L}_{0,1})$, respectively. We also assume
that the solution $\Phi_n(x)$ depends smoothly on
$(\beta_1,...,\beta_N)$ in an open non-empty set of $\R^N$ and
introduce the Hessian matrix ${\cal U}$ with elements:
\begin{equation}
\label{Hessian} {\cal U}_{n,m} = \frac{\partial Q_n}{\partial
\beta_m},
\end{equation}
where $Q_n = Q_n(\beta_1,...,\beta_N) = \int_{\R} \Phi_n^2 dx$. We
denote the number of positive and zero eigenvalues of matrix
${\cal U}$ as $p({\cal U})$ and $z({\cal U})$, respectively.
Finally, we introduce the linearized Hamiltonian ("energy") of the
eigenvalues $\lambda$ in $H^1(\mathbb{R},\mathbb{C}^{2N})$:
\begin{equation}
\label{linHam} h[{\bf u},{\bf w}] = \langle {\bf u}, {\cal L}_1
{\bf u} \rangle + \langle {\bf w}, {\cal L}_0 {\bf w} \rangle,
\end{equation}
where $\langle \cdot, \cdot \rangle$ is the standard inner product
in $L^2(\mathbb{R},\mathbb{C}^{2N})$. The negative index of the
linearized Hamiltonian is the number of negative eigenvalues of
${\cal L}_1$ and ${\cal L}_0$ in $L^2(\R,\C^N)$.

Several assumptions are imposed on the linearized problem (\ref{L})
in a general case \cite{P03}: \\
\hspace{1cm} (i) $z({\cal L}_1) = 1$, $z({\cal L}_0) = N$; \\
\hspace{1cm} (ii) $z({\cal U}) = 0$; \\
\hspace{1cm} (iii) no eigenvalues $\lambda \in i \mathbb{R}$
exist with $h[{\bf u},{\bf w}] = 0$;\\
\hspace{1cm} (iv) no embedded eigenvalues $\lambda \in i \mathbb{R}$
exist with $|{\rm Im}(\lambda)| \geq \beta_{\rm min}$, where
$\beta_{\rm min} = \min(\beta_1,...,\beta_N).$

{\bf Closure Theorem} \cite{P03} {\em Assume that (i)--(iv) be
satisfied. Let $N_{\rm imag}^-$ be the number of eigenvalues
$\lambda \in i \mathbb{R}_+$ with $h[{\bf u},{\bf w}] < 0$. Then
\begin{eqnarray}
\label{closure} & \phantom{t} & (i) \quad N_{\rm real} + 2 N_{\rm
comp} + 2 N_{\rm imag}^- = n({\cal L}_1) - p({\cal U}) + n({\cal L}_0), \\
\label{unstabletheorem2} & \phantom{t} & (ii) \quad
N_{\rm real} \geq | n({\cal L}_1) - p({\cal U}) - n({\cal L}_0)|, \\
\label{unstabletheorem3} & \phantom{t} & (iii) \quad N_{\rm comp}
\leq \min \left( n({\cal L}_0), n({\cal L}_1) - p({\cal U})
\right),
\end{eqnarray}
such that
\begin{equation}
\label{unstabletheorem} |n({\cal L}_1) - p({\cal U}) - n({\cal
L}_0)| \leq N_{\rm unst} \leq n({\cal L}_1) - p({\cal U}) +
n({\cal L}_0),
\end{equation}
where $N_{\rm unst} = N_{\rm real} +  2 N_{\rm comp}$ is the total
number of unstable eigenvalues in the problem (\ref{L}). }

This theorem was originally proved for the coupled NLS equations
(\ref{NLS}) in one dimension \cite{P03} and then generalized to a
three-dimensional NLS equation \cite{CPV04} and to an abstract
Hamiltonian dynamical system \cite{KKS04}. It allows us to
analytically trace unstable eigenvalues under parameter
continuations, starting with the particular limits, where all
eigenvalues $\lambda$ of negative energy $h[{\bf u},{\bf w}]$ are
known. Examples of such parameter continuation are recently reported
in \cite{KK04,PY03} in the context of the coupled NLS equations.

Bifurcations of unstable eigenvalues may occur in the linearized
problem (\ref{L}), when operators ${\cal L}_1$ and ${\cal L}_0$
change according to a continuous deformation and one of the
assumptions (i)--(iv) of the Closure Theorem fails. Bifurcations are
reviewed in Appendix A. In what follows, we apply parameter
continuation and bifurcation analysis to the system of coupled NLS
equations (\ref{NLS}) with cubic and saturable nonlinearities.

\section{The coupled cubic NLS equations}

We consider the system of coupled cubic NLS equations
\cite{Benney,Hasegawa,Menyuk}:
\begin{eqnarray}
\nonumber  \label{CPNLS}
i \psi_{1z} + \psi_{1xx} + \left(
|\psi_1|^2 + \chi |\psi_2|^2 \right) \psi_1 & = & 0, \\
\label{2NLS} i \psi_{2z} + \psi_{2xx} + \left( \chi |\psi_1|^2 +
|\psi_2|^2 \right) \psi_2 & = & 0,
\end{eqnarray}
where $\chi > 0$. The system is a particular example of
(\ref{NLS}) with $N = 2$, $d_1 = d_2 = 1$, and
\begin{equation}
U = \frac{1}{2} |\psi_1|^4 + \chi |\psi_1|^2 |\psi_2|^2 +
\frac{1}{2} |\psi_2|^4.
\end{equation}
The system (\ref{2NLS}) has a countable infinite set of families of
vector solitons ${\bf \Phi}(x) = (\Phi_1,\Phi_2)^T$, classified by
different nodal index ${\bf i} = (i_1,i_2)^T$, where $i_n$ is the
number of zeros of $\Phi_n(x)$ on $x \in \R$
\cite{Champ,Haelterman,Yang}. We consider here families of vector
solitons with nodal index ${\bf i} = (0,n)^T$,  $n \in \mathbb{N}$,
which are locally close to the NLS soliton, ${\bf \Phi}_{\rm NLS}(x)
= (\Phi^{(0)},0)^T$, where $\Phi^{(0)}(x) = \sqrt{2 \beta_1} \; {\rm
sech}(\sqrt{\beta_1} x)$. We let $\beta_1 = 1$ and $\beta_2 = \beta$
for convenience and introduce scalar Schr\"{o}dinger operators:
\begin{eqnarray}
\label{Schr0}
L_0 & = & - \frac{d^2}{d x^2} + 1 - 2 \; {\rm sech}^2(x), \\
\label{Schr1}
L_1 & = & - \frac{d^2}{d x^2} + 1 - 6 \; {\rm sech}^2(x), \\
\label{SchrS} L_s & = & - \frac{d^2}{d x^2} + \lambda_n^2(\chi) -
2 \chi \; {\rm sech}^2(x),
\end{eqnarray}
where
\begin{equation}
\label{lambdan} \lambda_n(\chi) = \frac{\sqrt{1 + 8 \chi} - (2n +
1)}{2}.
\end{equation}
The scalar operators $L_0$,$L_1$,$L_s$ define the matrix operators
${\cal L}_0$ and ${\cal L}_1$ at $\epsilon = 0$:
$$
{\cal L}_0 = {\rm diag}(L_0,L_s), \qquad {\cal L}_1 = {\rm
diag}(L_1,L_s).
$$
We define the perturbation series expansions of vector solitons
${\bf \Phi} = (\Phi_1,\Phi_2)^T$:
\begin{eqnarray}
\nonumber \Phi_1(x) & = & \Phi^{(0)}(x)
+ \epsilon^2 \Phi^{(2)}(x) + {\rm O}(\epsilon^4),  \\
\label{series100} \Phi_2(x) & = & \epsilon \Phi^{(1)}(x) +
\epsilon^3 \Phi^{(3)}(x) + {\rm O}(\epsilon^5),
\end{eqnarray}
and
\begin{equation}
\label{series101} \beta = \lambda_n^2(\chi) + \epsilon^2 C_n(\chi) +
{\rm O}(\epsilon^4).
\end{equation}
Corrections of the perturbation series
(\ref{series100})--(\ref{series101}) satisfy the linear equations:
\begin{eqnarray}
\label{non1}
L_s \Phi^{(1)} & = & 0, \\
\label{non2} L_1 \Phi^{(2)} & = & \chi \Phi^{(0)} \left(
\Phi^{(1)} \right)^2, \\
 \label{non3} L_s \Phi^{(3)} & = &
-C_n(\chi) \Phi^{(1)} + 2 \chi \Phi^{(0)} \Phi^{(1)} \Phi^{(2)} +
\left( \Phi^{(1)} \right)^3.
\end{eqnarray}
The problem (\ref{non1}) has a decaying solution $\Phi^{(1)} \equiv
\Phi^{(1)}_n(x)$ (see \cite{Haelterman,Yang}). When $n = 0$ and $\chi >
0$, the solution $\Phi^{(1)}_0 = {\rm sech}^s(x)$, $s =
\lambda_0(\chi)$ is a ground state. When $n
> 0$ and $\chi > \chi_n = n(n+1)/2$, the solution $\Phi^{(1)}_n(x)$
is an excited state with exactly $n$ nodes on $x \in \R$. The
problem (\ref{non2}) also has a decaying solution $\Phi^{(2)}(x)$,
since the right-hand-side $\chi \Phi^{(0)} \left( \Phi^{(1)}_n
\right)^2$ is orthogonal to the kernel of the operator ${\cal L}_1$,
which is $\Phi^{(0) \prime}(x)$. By the Fredholm Alternative Theorem, the
problem (\ref{non3}) has a decaying solution if and only if the
right-hand-side is orthogonal to the kernel of ${\cal L}_s$, which
is $\Phi^{(1)}_n(x)$. The orthogonality condition defines the
parameter $C_n(\chi)$ in the form:
\begin{equation}
\label{Cpar} C_n(\chi) = \frac{\langle \left( \Phi^{(1)}_n
\right)^2, \left( 2 \chi \Phi^{(0)} \Phi^{(2)} + \left( \Phi^{(1)}_n
\right)^2 \right) \rangle}{\langle \Phi^{(1)}_n, \Phi^{(1)}_n
\rangle}.
\end{equation}
The condition $C_n(\chi) \neq 0$ gives the sufficient condition of
continuation of the perturbation series expansions
(\ref{series100})--(\ref{series101}). Thus, for $\chi > \chi_n$ and
$C_n(\chi) \neq 0$, there exists some $R_n > 0$, such that the
$n$-th family of vector solitons ${\bf \Phi}(x) = (\Phi_1,\Phi_2)^T$
with the nodal index ${\bf i} = (0,n)^T$ bifurcates from ${\bf
\Phi}_{\rm NLS} = (\Phi^{(0)},0)^T$ in the one-sided domain ${\cal
B}_n$:
\begin{equation}
\label{domain2} {\cal B}_n = \left\{ \beta : 0 < \left| \beta -
\lambda_n^2(\chi) \right| < R_n, \quad {\rm sign} \left( \beta -
\lambda_n^2(\chi) \right) = {\rm sign}(C_n(\chi)) \right\}.
\end{equation}
These results for the first three families $n=0,1,2$ were
analytically obtained and numerically verified in \cite{Yang}.
We investigate stability of the $n$-th family of vector solitons in
the one-sided domain ${\cal B}_n$ below.

\subsection{Analytical results}

We trace unstable eigenvalues using the Closure Theorem. We consider
a generic case $C_n(\chi) \neq 0$ in the one-sided open domain
${\cal B}_n$ and show that the left-hand and right-hand sides of the
closure relation (\ref{closure}) are equal to $2n$ for small
$\epsilon \geq 0$.

Operator $L_0$ in (\ref{Schr0}) has one bound state for zero
eigenvalue, operator $L_1$ in (\ref{Schr1}) has two bound states
for negative and zero eigenvalues, and operator $L_s$ in
(\ref{SchrS}) has $(n+1)$ bound states with $n$ negative and one
zero eigenvalues. Therefore, at $\epsilon = 0$, we have $n({\cal
L}_0) = 0 + n = n$, $z({\cal L}_0) = 1 + 1 = 2$, $n({\cal L}_1) =
1 + n$ and $z({\cal L}_1) = 1 + 1 = 2$. It follows from Sturm
Nodal Theorem that
$$
n({\cal L}_0) = n, \qquad z({\cal L}_0) = 2, \qquad \forall
\epsilon \geq 0.
$$
Since $z({\cal L}_1) = 2 > 1$, we have the bifurcation case $z({\cal
L}_1) > 1$ for $\epsilon = 0$ (see Appendix A.1). It is however a
degenerate bifurcation case, since it occurs on the boundary of the
existence domain ${\cal B}_n$, such that $\beta \in \partial {\cal
B}_n$. We trace the zero eigenvalue of ${\cal L}_1$ for $\epsilon
\neq 0$ by the regular perturbation series,
\begin{eqnarray}
\label{pertur1}
{\bf u}(x) = \left[ \begin{array}{c} 0 \\
\Phi^{(1)}_n(x) \end{array} \right] + \epsilon \left[
\begin{array}{c} u^{(1)}(x) \\ 0 \end{array} \right]
+ \epsilon^2 \left[ \begin{array}{c} 0 \\
u^{(2)}(x) \end{array} \right] + {\rm O}(\epsilon^3)
\end{eqnarray}
and
\begin{eqnarray}
\lambda = \epsilon^2 \lambda_2 + {\rm O}(\epsilon^4).
\end{eqnarray}
Corrections of the perturbation series (\ref{pertur1}) satisfy a
set of linear non-homogeneous equations:
\begin{eqnarray}
\label{non4}
L_1 u^{(1)} & = & 2 \chi \Phi^{(0)}
\left( \Phi^{(1)}_n \right)^2, \\
\label{non5} L_s u^{(2)} & = & (\lambda_2 - C_n(\chi))
\Phi^{(1)}_n + 2 \chi \Phi^{(0)} \Phi^{(1)}_n \left( u^{(1)} +
\Phi^{(2)} \right) + 3 \left( \Phi^{(1)}_n \right)^3.
\end{eqnarray}
It follows from (\ref{non2}) and (\ref{non4}) that $u^{(1)} = 2
\Phi^{(2)}$. By the Fredholm Alternative Theorem, decaying solutions of
(\ref{non5}) exist if and only if the right-hand-side of
(\ref{non5}) is orthogonal to $\Phi^{(1)}_n(x)$. Using
(\ref{Cpar}), we find that $\lambda_2 = - 2 C_n(\chi)$. Therefore,
we have:
$$
n({\cal L}_1) = 1 + \Theta(C_n(\chi)) + n, \qquad z({\cal L}_1) =
1, \qquad \epsilon > 0,
$$
where $\Theta(z)$ is the Heaviside step-function. We trace the zero
eigenvalue of ${\cal U}$ from (\ref{Hessian}) and (\ref{series100}):
\begin{eqnarray*}
{\cal U}_{1,1} & = & \frac{\partial Q_1}{\partial \beta_1}
\biggr|_{\beta_1 = 1} = 2 + 2 \langle \Phi^{(0)}, \Phi^{(2)}
\rangle \frac{\partial \epsilon^2}{\partial \beta_1}
\biggr|_{\beta_1 = 1} + {\rm O}(\epsilon^2), \\
{\cal U}_{1,2} & = & \frac{\partial Q_1}{\partial
\beta_2}\biggr|_{\beta_1 = 1} = 2 \langle \Phi^{(0)}, \Phi^{(2)}
\rangle \frac{\partial \epsilon^2}{\partial \beta_2}
\biggr|_{\beta_1 = 1} + {\rm O}(\epsilon^2), \\
{\cal U}_{2,1} & = & \frac{\partial Q_2}{\partial \beta_1}
\biggr|_{\beta_1 = 1} = \langle \Phi^{(1)}_n, \Phi^{(1)}_n \rangle
\frac{\partial \epsilon^2}{\partial \beta_1} \biggr|_{\beta_1 = 1}
+ {\rm O}(\epsilon^2), \\
{\cal U}_{2,2} & = & \frac{\partial Q_2}{\partial \beta_2}
\biggr|_{\beta_1 = 1} = \langle \Phi^{(1)}_n, \Phi^{(1)}_n \rangle
\frac{\partial \epsilon^2}{\partial \beta_2} \biggr|_{\beta_1 = 1}
+ {\rm O}(\epsilon^2).
\end{eqnarray*}
It follows from (\ref{series101}) that
\begin{equation}
\frac{\partial \epsilon^2}{\partial \beta_2} \biggr|_{\beta_1 = 1}
= \frac{1}{C_n(\chi)} + {\rm O}(\epsilon^2)
\end{equation}
and, due to the symmetry of ${\cal U}$,
\begin{equation}
\label{detU-hessian} \det({\cal U}) = \frac{2 \langle
\Phi^{(1)}_n, \Phi^{(1)}_n \rangle}{C_n(\chi)} + {\rm
O}(\epsilon^2).
\end{equation}
Therefore, we have:
$$
p({\cal U}) = 1 + \Theta(C_n(\chi)), \qquad z({\cal U}) = 0,
\qquad \epsilon > 0.
$$
We conclude that the bifurcation case $z({\cal L}_1) > 1$ on the
boundary of the existence domain $\beta \in \partial {\cal B}_n$
does not result in bifurcation of any eigenvalue $\lambda$ of the
stability problem (\ref{L}), such that $n({\cal L}_1) - p({\cal U})
+ n({\cal L}_0) = 2n$ is valid everywhere in $\beta \in {\cal B}_n
\cup \partial {\cal B}_n$. It follows from the Closure Theorem that
the ground state with $n = 0$ is spectrally stable in $\beta \in
{\cal B}_n$, while the $n$-th excited state with $n \geq 1$ may have
at most $N_{\rm unst}$ unstable eigenvalues, where $0 \leq N_{\rm
unst} \leq 2n$. We show that $N_{\rm unst} = 2 N_{\rm comp} = 2 n$
in $\beta \in {\cal B}_n$ in a generic case.

At $\epsilon = 0$, the stability problem (\ref{L}) can be
decoupled as follows:
\begin{equation}
\label{problem1} L_1 u_1 = - \lambda w_1, \qquad L_0 w_1 = \lambda
u_1
\end{equation}
and
\begin{equation}
\label{problem2} L_s (u_2 \pm i w_2) = \pm i\lambda (u_2 \pm i
w_2).
\end{equation}
The first problem (\ref{problem1}) has the continuous spectrum for
${\rm Re}(\lambda) = 0$ and $|{\rm Im}(\lambda)| \geq 1$ and the
zero eigenvalue $\lambda = 0$ of algebraic multiplicity 4 and
geometric multiplicity 2. The second problem
(\ref{problem2}) has the continuous spectrum for ${\rm
Re}(\lambda) = 0$ and $|{\rm Im}(\lambda)| \geq
\lambda_n^2(\chi)$, zero eigenvalue $\lambda = 0$ of geometric and
algebraic multiplicity 2, and $2n$ isolated eigenvalues in the
points $\lambda = \pm i \left( \lambda_k^2 - \lambda_n^2 \right)$,
where $k = 0,1,...,n-1$. It follows from (\ref{lambdan}) that for
$\chi > \chi_n$:
\begin{equation}
\label{range-eigenvalue} \lambda_k^2 - \lambda_n^2 = (n-k) \left[ 2
\lambda_n(\chi) + (n-k) \right] > (n-k)^2 \geq 1, \qquad 0 \leq k <
n.
\end{equation}
Therefore, $2n$ isolated eigenvalues of the problem
(\ref{problem2}) are embedded in the continuous spectrum of the
problem (\ref{problem1}). These embedded eigenvalues have negative
energy $h[{\bf u},{\bf w}]$, since at $\epsilon = 0$:
\begin{equation}
\langle {\bf u}_k, {\cal L}_1 {\bf u}_k \rangle = \langle {\bf w}_k,
{\cal L}_0 {\bf w}_k \rangle = - (\lambda_k^2 - \lambda_n^2) \langle
\Phi_k^{(1)}, \Phi_k^{(1)} \rangle, \qquad 0 \leq k < n,
\end{equation}
where ${\bf u}_k = (0, \Phi_k^{(1)})^T$ and ${\bf w}_k = (0,\mp i
\Phi_k^{(1)})^T$ at $\epsilon = 0$. By Appendix A.4, all $2n$
embedded eigenvalues of negative energy $h[{\bf u},{\bf w}]$
bifurcate in a general case of non-zero $\Gamma$, see Eq.
(\ref{Fermat}), to complex unstable eigenvalues $\lambda \in \C$,
${\rm Re}(\lambda) > 0$ for $\epsilon \neq 0$, such that $N_{\rm
real} + 2 N_{\rm comp} + 2 N_{\rm imag}^- = 2 N_{\rm comp} = 2n$ in
$\beta \in {\cal B}_n$.

\subsection{Numerical results}

For $\epsilon \neq 0$, the linearized problem (\ref{L}) satisfies
the assumptions of the Closure Theorem. Therefore, all unstable
eigenvalues $N_{\rm unst} = 2 N_{\rm comp} = 2 n$ are structurally
stable for larger values of $\epsilon$, until new bifurcations occur
in the parameter continuations. We study numerically locations of
unstable eigenvalues in the linearized problem (\ref{L}) related to
the vector solitons ${\bf \Phi} = (\Phi_1,\Phi_2)^T$ with nodal
index ${\bf i} = (0,n)^T$, $n = 1,2$. Our numerical algorithm is
based on the shooting technique in the complex $\lambda$-plane. We
also determine the indices $n({\cal L}_0)$, $n({\cal L}_1)$ and
$p({\cal U})$ by a numerics-assisted procedure as described in
\cite{YangPRE02}, and relate them to the number of unstable
eigenvalues by using the closure relation (\ref{closure}).

Figure 1 shows the 1st-family of multi-hump vector solitons with the
correspondence: $u = \Phi_1(x)$, $v = \Phi_2(x)$, and $\omega =
\sqrt{\beta}$. For $\omega < 1$, this family exists between
$\chi_1(\beta) < \chi < \chi_2(\beta)$, where $\chi=\chi_2(\beta)$
is the local bifurcation boundary, and $\chi=\chi_1(\beta)$ is the
nonlocal bifurcation boundary \cite{Champ}. Hence the one-sided
domain $\beta \in {\cal B}_1$ is located to the left of the local
bifurcation curve, and ${\rm sign}(C_1) = 1$ in (\ref{domain2}).
When parameter $\omega = 0.6$ is fixed, we readily find that $\chi_1
=0.28$ and $\chi_2 = 2.08$. When $\chi$ moves from $\chi_2$ to
$\chi_1$, the distance between the two pulses in the $v$ component
grows. It diverges to infinity at the nonlocal bifurcation boundary
$\chi_2$ (near point $a$ in Fig. 1).

Figure 2 shows unstable eigenvalues of the linearized problem
(\ref{eigenvalue-problem}) for $\omega = 0.6$ and $\chi_1 < \chi <
\chi_2$. In the domain $\beta \in {\cal B}_1$, there is a pair of
unstable complex eigenvalues $\lambda = {\rm Re}(\sigma_2) \pm i
{\rm Im}(\sigma_2)$, which bifurcate from the embedded eigenvalues
$\lambda = \pm i (\lambda_0^2 - \lambda_1^2)$, see Eq.
(\ref{range-eigenvalue}).

When $\chi \to 1^+$, the complex eigenvalues $\lambda = {\rm
Re}(\sigma_2) \pm i {\rm Im}(\sigma_2)$ approach the imaginary
axis and become embedded eigenvalues $\lambda = \pm i (1 -
\omega^2)$. The case $\chi = 1$ corresponds to the integrable
Manakov system, when the linearized problem
(\ref{eigenvalue-problem}) has the following exact solution:
\begin{equation}
\label{particularsolution}
{\bf u}_0 = \left( \begin{array}{cc} -\Phi_2 \\
\Phi_1 \end{array} \right) \qquad
{\bf w}_0 = \mp i \left( \begin{array}{cc} \Phi_2 \\
\Phi_1 \end{array} \right), \qquad \lambda = \pm i (1 - \omega^2).
\end{equation}
This exact solution is generated by the additional
polarizational-rotation symmetry in the potential function $U =
U(|\psi_1|^2 + |\psi_2|^2)$ at $\chi = 1$. Embedded eigenvalues
$\lambda = \pm i (1 - \omega^2)$ have negative energy $h[{\bf
u},{\bf w}]$, since
\begin{equation}
\label{energyquadraticform} \langle {\bf u}_0, {\cal L}_1 {\bf
u}_0 \rangle = \langle {\bf w}_0, {\cal L}_0 {\bf w}_0 \rangle = -
(1 - \omega^2) \left( \langle \Phi_1, \Phi_1 \rangle - \langle
\Phi_2, \Phi_2 \rangle \right) < 0,
\end{equation}
where the last inequality is confirmed numerically from
integration of the exact solutions \cite{Tratnik}:
\begin{eqnarray}
\Phi_1(x) & = & \frac{\sqrt{1 - \omega^2} \cosh \omega x}{\cosh x
\cosh \omega x - \omega \sinh x \sinh \omega x}, \\
\Phi_2(x) & = & -\frac{\omega \sqrt{1 - \omega^2} \sinh x}{\cosh x
\cosh \omega x - \omega \sinh x \sinh \omega x},
\end{eqnarray}
in the entire domain of existence: $0 < \omega < 1$. Since embedded
eigenvalues $\lambda = \pm i (1 - \omega^2)$ at $\chi = 1$ have
negative energy $h[{\bf u},{\bf w}]$, they bifurcate to the complex
plane for $\chi \neq 1$ when the polarizational symmetry is
destroyed. This is indeed the case as shown in Fig. 2, in agreement
with Appendix A.4.

There is an additional instability bifurcation at $\chi = 1$. This
bifurcation comes about because at this $\chi$ value, the 1st-family
of vector solitons ${\bf \Phi} = (\Phi_1,\Phi_2)^T$ can be
generalized to asymmetric solutions with an additional free
parameter \cite{Akhmediev,Haelterman,Yang}. As a result, the
derivative of the asymmetric vector solitons with respect to the
free parameter is an eigenvector in the kernel of the operator
${\cal L}_1$, such that $z({\cal L}_1) = 2$ at $\chi = 1$. When
$\chi \neq 1$, the integrability of the Manakov system is destroyed,
and a pair of real or purely imaginary eigenvalues is generated, in
agreement with Appendix A.1. Indeed, Fig. 2 shows a pair of purely
imaginary eigenvalues $\lambda = \pm i \sigma_1$ for $\chi > 1$,
which merge to the end points $\lambda = \pm i \omega^2$ of the
continuous spectrum at $\chi = 1.185$, and a pair of real
eigenvalues $\lambda = \pm \sigma_1$ for $\chi < 1$.

Now we relate results of Fig. 2 to the closure relation
(\ref{closure}). For this purpose, we have determined the indices
$n({\cal L}_1)$ and $p({\cal U})$ by the numerics-assisted
procedure in \cite{YangPRE02} (we note that $n({\cal L}_0) = 1$
everywhere in the existence domain of the 1st-family of vector
solitons). For $\omega=0.6$, we have found numerically that
\begin{equation}
n({\cal L}_1) = \left\{
\begin{array}{c} 4, \hspace{0.3cm} \chi_1 < \chi < 1, \\
3, \hspace{0.3cm} 1< \chi < \chi_2, \end{array} \right. \qquad
p({\cal U}) = 2, \; \mbox{for all $\chi_1<\chi<\chi_2$},
\end{equation}
such that
\begin{equation}
n({\cal L}_1) + n({\cal L}_0) - p({\cal U}) = \left\{
\begin{array}{c} 3, \hspace{0.3cm} \chi_1 < \chi < 1, \\
                 2, \hspace{0.3cm} 1 < \chi < \chi_2. \end{array} \right.
\end{equation}
On the other hand, Fig. 2 shows that
\begin{equation}
N_{\rm comp} = 1, \; \mbox{for all $\chi_1<\chi<\chi_2$}, \qquad
N_{\rm real} = \left\{ \begin{array}{c} 1, \hspace{0.3cm} \chi_1<\chi < 1 \\
0, \hspace{0.3cm} 1< \chi < \chi_2 \end{array} \right.
\end{equation}
and the closure relation (\ref{closure}) is thus satisfied.

Figures 3 and 4 show similar results for the 2nd family of
multi-hump vector solitons. When parameter $\omega=0.6$ is fixed,
the local bifurcation boundary is $\chi_2 = 4.68$ and the
nonlocal bifurcation boundary is $\chi_1 = 1.68$. Again, the
one-sided domain $\beta \in {\cal B}_2$ is located to the left of
the local bifurcation boundary, such that ${\rm sign}(C_2) = 1$.
However, such solitons exist on {\em both}
sides of the nonlocal bifurcation boundary $\chi=\chi_1$. As
$\chi$ moves leftward from $\chi=\chi_2$, it first crosses the
nonlocal bifurcation boundary $\chi=\chi_1$, then turns around,
and approaches the nonlocal bifurcation boundary from the {\em
left} side (see point $a$ in Fig. 3). This behavior has been explained
both analytically and numerically in \cite{Champ}.

Using the same shooting algorithm, we have obtained the unstable
eigenvalues of the linearized problem (\ref{L}) and displayed them
in Fig. 4. In the domain $\beta \in {\cal B}_2$, there exist two
pairs of unstable complex eigenvalues, such that the pair $\lambda =
{\rm Re}(\sigma_4) \pm i {\rm Im}(\sigma_4)$ bifurcates from the
pair of embedded eigenvalues $\lambda = \pm i( \lambda_0^2 -
\lambda_2^2)$, while the pair $\lambda = {\rm Re}(\sigma_3) \pm i
{\rm Im}(\sigma_3)$ bifurcates from the pair of embedded eigenvalues
$\lambda = \pm i ( \lambda_1^2 - \lambda_2^2)$. Fig. 4 also shows
that the pair $\lambda = {\rm Re}(\sigma_3) \pm i {\rm
Im}(\sigma_3)$ approach the imaginary axis and become a pair of
embedded eigenvalues at $\chi = \chi_a \approx 2.49$, but then
reappear as a pair of complex unstable eigenvalues for $\chi <
\chi_a$, in agreement with Appendix A.4.

There are two more instability bifurcations in Fig. 4. At
$\chi = \chi_b \approx 2.44$, the
zero eigenvalue bifurcates into a pair of imaginary
eigenvalues $\lambda = \pm i \sigma_2$ for $\chi
> \chi_b$, which then merge into the end points $\lambda = \pm i
\omega^2$ of the continuous spectrum at $\chi \approx 2.72$.
When $\chi < \chi_b$, this zero eigenvalue bifurcates into a pair of real unstable
eigenvalues $\lambda = \pm \sigma_2$.
At yet another point $\chi=\chi_c=2.17$, the zero eigenvalue
bifurcates into a pair of imaginary
eigenvalues $\lambda = \pm i \sigma_1$ for $\chi > \chi_c$, which merge
into the end points $\lambda = \pm i \omega^2$ of the continuous spectrum
at $\chi \approx 3.01$.
When $\chi < \chi_c$, this zero eigenvalue bifurcates into
a pair of real unstable eigenvalues $\lambda = \pm \sigma_1$.

To relate numerical results of Fig. 4 to the closure relation
(\ref{closure}), we have again determined the indices $n({\cal
L}_1)$ and $p({\cal U})$ by the numerical algorithm, while
$n({\cal L}_0) = 2$ everywhere in the existence domain of the 2nd
family of vector solitons. For $\omega=0.6$, we have found
numerically that
\begin{equation}
n({\cal L}_1) = \left\{\begin{array}{c} 5, \hspace{0.3cm} \chi_1 < \chi < \chi_b, \\
4, \hspace{0.3cm} \chi_b < \chi < \chi_2, \end{array} \right.
\qquad p({\cal U}) = \left\{\begin{array}{c} 1, \hspace{0.3cm} \chi_1 < \chi < \chi_c, \\
2, \hspace{0.3cm} \chi_c < \chi < \chi_2, \end{array} \right.
\end{equation}
such that
\begin{equation}
n({\cal L}_1)+n({\cal L}_0)-p({\cal U}) = \left\{
\begin{array}{l} 6, \hspace{0.3cm} \chi_1 < \chi < \chi_c, \\
                 5, \hspace{0.3cm} \chi_c < \chi < \chi_b, \\
                 4, \hspace{0.3cm} \chi_b < \chi < \chi_2.
                 \end{array} \right.
\end{equation}
On the other hand, Fig. 4 shows that
\begin{equation}
N_{\rm comp} = 2, \; \mbox{for all $\chi_1 < \chi < \chi_2$}, \qquad
N_{\rm real} = \left\{\begin{array}{l} 2, \hspace{0.3cm} \chi_1 < \chi < \chi_c, \\
                                   1, \hspace{0.3cm} \chi_c < \chi < \chi_b, \\
                                   0, \hspace{0.3cm} \chi_b < \chi < \chi_2.
                                   \end{array} \right.
\end{equation}
Hence the closure relation (\ref{closure}) is satisfied. In
particular, the instability bifurcation at $\chi=\chi_b$ is due to a
jump in the index $n({\cal L}_1)$ (similar to the 1st family), while
the instability bifurcation at $\chi=\chi_c$ is due to a jump in the
index $p({\cal U})$. These two bifurcations occur in agreement with
Appendices A.1 and A.2.

Although our results were obtained here for a particular value
$\omega=0.6$, we expect that similar results hold for other values
of $\omega$, when $0 < \omega < 1$. We conclude that the 1st and
2nd families of multi-hump vector solitons in the coupled cubic
NLS equations are all linearly unstable (except for the integrable
Manakov system $\chi=1$, where the 1st family is neutrally
stable).

It is remarkable that the main features of instability
bifurcations for the 1st-family of multi-hump vector solitons are
repeated for the 2nd-family of vector solitons, irrelevant whether
the coupled NLS equations are integrable or not. This allows us to
conjecture that a similar pattern of unstable eigenvalues persists
for a general $n$-th family of multi-hump vector solitons, with
more unstable eigenvalues and additional instability bifurcations
appearing as $n$ increases.

\section{Two coupled saturable NLS equations}

We consider the system of two coupled saturable NLS equations
\cite{Chris_Segev96,Kivshar_book}:
\begin{eqnarray}   \label{Saturable}
\nonumber i \psi_{1z} + \psi_{1xx} + \frac{|\psi_1|^2 +
|\psi_2|^2}{1 + s(|\psi_1|^2 + |\psi_2|^2)} \psi_1 & = & 0, \\
\label{3NLS} i \psi_{2z} + \psi_{2xx} + \frac{|\psi_1|^2 +
|\psi_2|^2}{1 + s(|\psi_1|^2 + |\psi_2|^2)} \psi_2  & = & 0,
\end{eqnarray}
where $s > 0$. This system is a particular example of (\ref{NLS})
with $N = 2$, $d_1 = d_2 = 1$, and
\begin{equation}
U = \frac{1}{s} \left( |\psi_1|^2 + |\psi_2|^2 \right) -
\frac{1}{s^2} \log \left( 1 + s (|\psi_1|^2 + |\psi_2|^2) \right).
\end{equation}

We consider again the $n$-th family of vector solitons ${\bf \Phi}
= (\Phi_1,\Phi_2)^T$ with nodal index ${\bf i} = (0,n)^T$, $n \in
\mathbb{N}$ and convenient parametrization $\beta_1 = 1$ and
$\beta_2 = \beta$. The $n$-th family bifurcates from the scalar
solution ${\bf \Phi} = (\Phi_0,0)^T$, where $\Phi_0(x)$ satisfies
the ODE:
\begin{equation}
\Phi_0'' - \Phi_0 + \frac{\Phi_0^3}{1 + s \Phi_0^2} = 0.
\end{equation}
The local bifurcation occurs at $\beta = \beta_n(s)$, when there
exists a $n$-nodal bound state $\Phi_n(x)$ in the linear eigenvalue
problem:
\begin{equation}
\label{bound-state-problem} \Phi_n'' - \beta_n \Phi_n +
\frac{\Phi_0^2 \Phi_n}{1 + s \Phi_0^2} = 0.
\end{equation}
Vector solitons in the $n$-th family disappear at the nonlocal
bifurcation boundary. The domain of existence for the first three
families $n = 0,1,2$ has been obtained numerically in
\cite{Champ,KivSkryabin}. We trace analytically unstable eigenvalues
and show that $N_{\rm real} + 2 N_{\rm comp} + 2 N_{\rm imag}^-  = 2
n$ for small $|\beta - \beta_n(s)| \ll 1$. At $\beta = \beta_n(s)$,
the stability problem (\ref{L}) can be decoupled as follows:
\begin{equation}
\label{problem21} L_1 u_1 = - \lambda w_1, \qquad L_0 w_1 =
\lambda u_1
\end{equation}
and
\begin{equation}
\label{problem22} L_s (u_2 \pm i w_2) = \pm i\lambda (u_2 \pm i
w_2),
\end{equation}
where
\begin{eqnarray}
L_0 & = & - \frac{d^2}{d x^2} + 1 - \frac{\Phi_0^2}{1 + s \Phi_0^2}, \\
L_1 & = & - \frac{d^2}{d x^2} + 1 - \frac{\Phi_0^2 (3 + s
\Phi_0^2)}{(1 + s \Phi_0^2)^2}, \\ \label{Lsss} L_s & = & -
\frac{d^2}{d x^2} + \beta_n(s) - \frac{\Phi_0^2}{1 + s \Phi_0^2}.
\end{eqnarray}
The first problem (\ref{problem21}) is the linearized stability
problem in the scalar saturable NLS equation (\ref{3NLS}) for
$\Phi_0(x)$. Based on numerical data in
\cite{KivOstrov,KivSkryabin}, we assume that the bound state
$\Phi_0(x)$ is spectrally stable in the scalar saturable NLS
equation and the problem (\ref{problem21}) does not have any
eigenvalues of negative energy $h[{\bf u},{\bf w}]$. It has the
continuous spectrum at ${\rm Re}(\lambda) = 0$ and $|{\rm
Im}(\lambda)| \geq 1$, the zero eigenvalue $\lambda = 0$ of
algebraic multiplicity 4 and geometric multiplicity 2, and possibly
isolated eigenvalues $\lambda \in i \R$ of positive energy.

The second problem (\ref{problem22}) has the continuous spectrum at
${\rm Re}(\lambda) = 0$ and $|{\rm Im}(\lambda)| \geq \beta_n(s)$,
zero eigenvalue $\lambda = 0$ of geometric and algebraic
multiplicity 2,  and $2n$ isolated eigenvalues $\lambda = \pm i [
\beta_k(s) - \beta_n(s) ]$, where $k = 0,1,...,n-1$. By the Sturm
Nodal Theorem, eigenvalues $\beta_k(s)$ ($k = 0,...,n$) are ordered
in the decreasing order and are characterized by eigenfunctions
$\Phi_k(x)$ with $k$ nodes, such that the ground state $\Phi_0(x)$
corresponds to $\beta_0(s) = 1$ and the $n$-th excited state
$\Phi_n(x)$ corresponds to $\beta_n(s)$. The $2n$ eigenvalues have
negative energy $h[{\bf u},{\bf w}]$, since
\begin{equation} \label{hn}
\langle {\bf u}_k, {\cal L}_1 {\bf u}_k \rangle = \langle {\bf w}_k,
{\cal L}_0 {\bf w}_k \rangle = - \left[ \beta_k(s) - \beta_n(s)
\right] \langle \Phi_k, \Phi_k \rangle < 0, \qquad 0 \leq k < n,
\end{equation}
where ${\bf u}_k = (0, \Phi_k)^T$ and ${\bf w}_k = (0,\mp i
\Phi_k)^T$ at $\beta = \beta_n(s)$. Therefore, $2 N_{\rm imag}^- = 2
n$ at $\beta = \beta_n(s)$. Since the zero eigenvalue has the
generic algebraic multiplicity six, no negative eigenvalues of
${\cal L}_1$ and ${\cal L}_0$ arise from the zero eigenvalue, such
that we have $N_{\rm real} + 2 N_{\rm comp} + 2 N_{\rm imag}^- = 2n$
for $|\beta-\beta_n(s)| \ll 1$ by continuity of the negative index
$n(h)$. The ground state with $n = 0$ is therefore spectrally stable
in the existence domain (which is $\beta = 1$, $s
> 0$ for $n = 0$).

We show that the nodal bound state with $n \geq 1$ may have at most
$N_{\rm unst}$ unstable eigenvalues, where $0 \leq N_{\rm unst} \leq
2n-2$. The number of unstable eigenvalues is reduced by two, since
there are two eigenvalues $\lambda = \pm i (1 - \beta)$ of negative
energy, which exist for all vector solitons with $n \geq 1$ due to
the polarizational-rotation symmetry in the potential function $U =
U(|\psi_1|^2 + |\psi_2|^2)$. This pair of eigenvalues is similar to
the one which occurs in the coupled NLS equations (\ref{2NLS}) with
$\chi = 1$, such that the stability problem (\ref{L}) has exactly
the same solution (\ref{particularsolution}) for $\beta = \omega^2$.
These eigenvalues have negative energy $h[{\bf u},{\bf w}]$ due to
(\ref{energyquadraticform}), where the inequality remains true in
the entire existence domain, as follows from numerical data in
\cite{KivSkryabin}. If $\beta < 1/2$, these eigenvalues $\lambda =
\pm i (1 - \beta)$ are embedded in the continuous spectrum of the
problem (\ref{L}), but never bifurcate off the continuous spectrum
as the parameters vary. Therefore, $2 N_{\rm imag}^- \geq 2$ and $0
\leq N_{\rm unst} \leq 2n-2$ for $n \geq 1$. As a result, the 1st
family of vector solitons is spectrally stable when $\beta$ is near
the local bifurcation boundary $\beta=\beta_n(s)$.

When $\beta = \beta_2(s)$, the eigenvalues $\lambda = \pm i \left[
\beta_1(s) - \beta_2(s) \right]$ are embedded into the continuous
spectrum if $\beta_1(s) > 2 \beta_2(s)$, and are isolated from the
continuous spectrum if $\beta_1(s) < 2 \beta_2(s)$. In the first
case, the embedded eigenvalues bifurcate generally to complex
unstable eigenvalues for $0<|\beta -\beta_2(s)|\ll 1$, according to
Appendix A.4. In the second case, the isolated eigenvalues do not
bifurcate to complex unstable eigenvalues for $0<|\beta
-\beta_2(s)|\ll 1$. Since these eigenvalues have negative energy
[see Eq. (\ref{hn})], the number of unstable eigenvalues $N_{\rm
unst}$ is then zero. In other words, vector solitons of the 2nd
family near the local bifurcation boundary with $\beta_1(s) < 2
\beta_2(s)$ are spectrally stable.

We compare the above theoretical results with numerical results in
\cite{KivSkryabin}, where unstable eigenvalues in the linearized
problem (\ref{L}) were obtained for the 1st and 2nd families [see
Figs. 2 and 3 in \cite{KivSkryabin}, where $(\lambda, \beta)$
correspond to our parameters $(\beta, \sigma)$]. It was shown in
\cite{KivSkryabin} that the 1st-family of vector solitons is stable
in the domain $\beta_1(s) < \beta < \beta_{\rm stab}^{(1)}(s)$, in
agreement with our prediction of linear stability near the first
local bifurcation boundary $\beta = \beta_1(s)$. At $\beta =
\beta_{\rm stab}^{(1)}(s)$, the bifurcation $z({\cal U}) = 1$
occurs, which generates a pair of real eigenvalues for $\beta >
\beta_{\rm stab}^{(1)}(s)$ and a pair of imaginary eigenvalues for
$\beta < \beta_{\rm stab}^{(1)}(s)$, in agreement with Appendix A.2.
Therefore,
\begin{equation}
n({\cal L}_1)+n({\cal L}_0)-p({\cal U}) = \left\{
\begin{array}{l} 3, \hspace{0.3cm} \beta_{\rm stab}^{(1)} < \beta < 1, \\
                 2, \hspace{0.3cm} \beta_1 < \beta < \beta_{\rm
                 stab}^{(1)},
                 \end{array} \right.
\end{equation}
and
\begin{equation}
N_{\rm imag}^- = 2, \; \mbox{for all $\beta$}, \qquad
N_{\rm real} = \left\{\begin{array}{l}
    1, \hspace{0.3cm} \beta_{\rm stab}^{(1)} < \beta < 1, \\
    0, \hspace{0.3cm} \beta_1 < \beta < \beta_{\rm stab}^{(1)}.
                 \end{array} \right.
\end{equation}

For the 2nd family, the eigenvalues $\lambda = \pm i \left[
\beta_1(s) - \beta_2(s) \right]$ are embedded when $0.646<s<0.857$
and isolated when $s > 0.857$. According to our prediction, the
embedded eigenvalues should bifurcate to the complex unstable
eigenvalues for $0<|\beta -\beta_2(s)|\ll 1$. However, it was
claimed in \cite{KivSkryabin} that vector solitons in the 2nd family
were stable near the local bifurcation boundary $0< |\beta -
\beta_2(s) | \ll 1$ for all values of $s$. In order to resolve this
discrepancy, we have numerically determined the eigenvalue spectrum
for vector solitons in the 2nd family by the shooting method.
Figures 5 and 6 present numerical results for $\beta = 0.16$ and
$\beta = 0.49$, respectively.

For $\beta=0.16$, the local bifurcation boundary of the 2nd family
occurs at $s=0.785$ and the eigenvalues $\lambda = \pm i \left[
\beta_1(s) - \beta_2(s) \right]$ are embedded, such that a pair of
unstable complex eigenvalues $\lambda = {\rm Re}(\sigma_2) \pm i
{\rm Im}(\sigma_2)$ bifurcates for $s < 0.785$. These complex
eigenvalues indicate the {\em oscillatory instability} of vector
solitons near the local bifurcation boundary $|\beta - \beta_2(s)|
\ll 1$ for $\beta = 0.16$. Thus the claim in \cite{KivSkryabin} on
the stability of 2nd-family vector solitons anywhere near the local
bifurcation boundary is incorrect. Note that the complex eigenvalues
$\lambda = {\rm Re}(\sigma_2) \pm i {\rm Im}(\sigma_2)$ persist
throughout the entire existence domain of the 2nd family, which is
$0.513 < s < 0.785$ at $\beta=0.16$.

Furthermore, a pair of purely imaginary eigenvalues $\lambda = \pm i
|\sigma_1|$ bifurcate from the end points $\lambda = \pm i \beta$ of
the continuous spectrum at $s=0.763$, merge into the origin at
$s=0.696$, and then bifurcate into a pair of real eigenvalues
$\lambda = \pm \sigma_1$ for $s < 0.696$. This {\em exponential
instability} induced by the real eigenvalue $\lambda = \sigma_1$ has
been reported in \cite{KivSkryabin}. Since it was claimed in
\cite{KivSkryabin} that $z({\cal U}) = 0$ in the entire existence
domain of the 2nd family of vector solitons, we conclude that the
instability bifurcation at $s = 0.696$ falls into the scenario of
Appendix A.1 with $z({\cal L}_1) = 2$. In the interval
$0.513<s<0.696$, the oscillatory instability is overshadowed by the
exponential instability from the eigenvalue $\lambda = \sigma_1$.
However, in the interval $0.696<s<0.785$, this oscillatory
instability is the only instability experienced by vector solitons.
Since ${\rm Re}(\sigma_2)$ is less than $0.011$ in the interval
$0.696<s<0.785$, it may explain why this oscillatory instability was
missed in the numerical results of \cite{KivSkryabin}.

For $\beta=0.49$, the local bifurcation boundary of the 2nd family
occurs at $s=0.894$, and the eigenvalues $\lambda = \pm i \left[
\beta_1(s) - \beta_2(s) \right]$ are isolated, such that vector
solitons near the local bifurcation boundary are spectrally stable.
This is indeed confirmed in Fig. 6, where the spectrum diagram is
displayed for all values of $s$. It is seen that at the local
bifurcation boundary $s=0.894$, there are two pairs of isolated
imaginary eigenvalues. The pair $\lambda = \pm i \left[ \beta_1(s) -
\beta_2(s) \right] = \pm 0.292i$ has the negative energy, while the
pair $\lambda = \pm 0.310i$ has the positive energy. The second pair
corresponds to the positive eigenvalue of ${\cal L}_s$ in
(\ref{Lsss}). As $s$ moves leftward from the boundary point 0.894,
these two imaginary eigenvalues move toward each other. At
$s=0.891$, they coalesce and create a quadruple of complex
eigenvalues, in agreement with Appendix A.3. However, this
instability persists only in a tiny interval $0.882<s<0.891$, and it
is very weak, with growth rates below 0.01. At $s=0.882$, these
complex eigenvalues coalesce and bifurcate back into two pairs of
purely imaginary eigenvalues again. When $s$ decreases further, one
pair of these imaginary eigenvalues (denoted as $\pm \sigma_1$ in
Fig. 6) always remain imaginary, but the other pair of imaginary
eigenvalues ($\pm \sigma_2$ in Fig. 6) move toward zero and become
real for $s<0.80$, in agreement with Appendix A.1. There is one more
eigenvalue ($\sigma_3$) in Fig. 6, which bifurcates from the edge of
the continuous spectrum at $s=0.718$, and always stays imaginary.
The pattern of Fig. 6 differs from that of Fig. 5 in that complex
instability is not set in at the local bifurcation boundary $\beta =
\beta_2(s)$, and, once it is set in, it is confined in a narrow
interval of $s$. Similar to the $\beta=0.16$ case above, this narrow
interval of oscillatory instability was missed in
\cite{KivSkryabin}, but the exponential instability was captured
there.

Finally, we have mapped out the regions of exponential and
oscillatory instabilities in the entire domain of existence for the
2nd family of vector solitons. The results are shown in Fig. 7. The
almost-straight boundary lines show the local and nonlocal
bifurcation boundaries \cite{Champ}. The large domain of exponential
instability away from the local bifurcation boundary corresponds to
the one computed in Fig. 2 of \cite{KivSkryabin}. The domain of
oscillatory instability consists of two sub-domains. The larger
sub-domain is at $s<0.857$, where the eigenvalues $\lambda = \pm i
\left[ \beta_1(s) - \beta_2(s) \right]$ at the local bifurcation
boundary $\beta = \beta_2(s)$ are embedded. The smaller sub-domain
is at $s>0.857$, where the eigenvalues $\lambda = \pm i \left[
\beta_1(s) - \beta_2(s) \right]$ are isolated but bifurcate to
complex eigenvalues away from the local bifurcation boundary $\beta
= \beta_2(s)$. Both sub-domains of oscillatory instabily were missed
in \cite{KivSkryabin}.

\section{Summary}

We have applied the Closure Theorem for the negative index of the
linearized Hamiltonian to multi-hump vector solitons in the general
coupled NLS equations (\ref{NLS}). Unstable eigenvalues of the
linearized problem (\ref{L}) are approximated with the perturbation
series expansions and found numerically with the shooting method.
Not only the numerical results are in excellent agreement with the
closure relation (\ref{closure}), but also the closure relation
(\ref{closure}) shows that all unstable eigenvalues are recovered
with the numerical shooting method. These analytical and numerical
results establish that all multi-hump vector solitons in the
non-integrable coupled {\em cubic} NLS equations are linearly
unstable, while multi-hump vector solitons in the coupled {\em
saturable} NLS equations can be linearly stable in certain regions
of the parameter space. In the latter case, we have also discovered
a new oscillatory instability which was missed before. This
oscillatory instability significantly reduces the stability domains
of vector solitons.

We note that the Closure Theorem is applied differently in Sections
3 and 4. For coupled cubic NLS equations, we compute the closure
relation (\ref{closure}) from the right-hand side $n({\cal L}_1) -
p({\cal U}) + n({\cal L}_0)$ at the local bifurcation boundary and
then match it with the number of unstable eigenvalues of the
linearized problem (\ref{L}). For coupled saturable NLS equations,
we compute the closure relation (\ref{closure}) directly from the
left-hand side $N_{\rm real} + 2 N_{\rm comp} + 2 N_{\rm imag}^-$ at
the local bifurcation boundary and continue it in the entire
existence domain.

While the negative index theory is well understood in \cite{P03} and
well illustrated in \cite{KKS04,KK04} and this paper, the following
problem still remains a challenge: {\em how do we understand the
stability of vector solitons in sign-indefinite coupled-mode
equations, such as the nonlinear Dirac equations?} The Closure
Theorem is obviously invalid for the Dirac equations as the
continuous spectrum has both positive and negative energies. Thus a
generalization of the Closure Theorem to such systems is highly
desirable for future advances.

{\bf Acknowledgements.} The work of D.E.P. was supported in part by
NSERC grant No. 5-36694. The work of J.Y. was supported in part by
NASA EPSCoR and AFOSR grants.

\appendix
\section{Appendix: Bifurcations of unstable eigenvalues}
\label{instability-bifurcation}

Here we classify the four special cases, when one of the assumptions
(i)--(iv) of the Closure Theorem fails. We derive sufficient
conditions, when new unstable eigenvalues with ${\rm Re}(\lambda) >
0$ bifurcate in the linearized problem (\ref{eigenvalue-problem})
from eigenvalues with ${\rm Re}(\lambda) = 0$. For clarity of
notations, we use an equivalent form of the spectral problem
(\ref{L}):
\begin{equation}
\label{diagonalizationproblem} {\cal L}_1 {\bf u} = \gamma {\cal
L}_0^{-1} {\bf u}, \qquad {\bf u} \in X_c^{(u)}(\R),
\end{equation}
where $\gamma = - \lambda^2$ and $X_c^{(u)}(\R)$ is the constrained
subspace of $L^2(\R)$:
\begin{eqnarray}
\label{constraint1} X_c^{(u)} = \left\{ {\bf u} \in L^2 : \; \langle
\Phi_n {\bf e}_n, {\bf u} \rangle = 0, \; n = 1,...,N \right\}.
\end{eqnarray}
Here ${\bf e}_1$,...,${\bf e}_N$ are unit vectors in $\R^N$ and none
of the components $\Phi_n(x)$ is assumed to vanish identically on $x
\in \R$. Operator ${\cal L}_0$ is always invertible in
$X_c^{(u)}(\R)$, since eigenvectors $\{ \Phi_n(x) {\bf e}_n
\}_{n=1}^N$ form a basis in the kernel of ${\cal L}_0$. In the
domain ${\cal D}_{\epsilon} = \left\{ \lambda \in \mathbb{C} : \;
|\lambda| > \epsilon \right\}$ for any $\epsilon
> 0$, there exists a relation between ${\bf u}(x)$ and ${\bf w}(x)$:
\begin{equation}
\label{w-u} {\bf w} = \lambda {\cal L}_0^{-1} {\bf u}, \qquad
\lambda \in {\cal D}_{\epsilon},
\end{equation}
such that two eigenvectors $({\bf u},{\bf w})^T$ and $({\bf u},-{\bf
w})^T$ of the linearized problem (\ref{L}) for $\lambda$ and
$(-\lambda)$ correspond to a single eigenvector ${\bf u}(x)$ of the
problem (\ref{diagonalizationproblem}) for $\gamma = - \lambda^2$.

\subsection{Zero eigenvalues of ${\cal L}_1$ and ${\cal L}_0$}

The kernel of ${\cal L}_0$ has a basis of $N$ eigenvectors $\{
\Phi_n(x) {\bf e}_n \}_{n=1}^N$. Therefore, the existence domain of
the vector solitons (\ref{soliton}) is confined by the boundaries,
where $\Phi_n(x) \equiv 0$ for some $n = 1,...,N$. Let ${\cal B}$ be
an open simply-connected domain in the parameter space
$(\beta_1,...,\beta_N)$, where the vector solitons (\ref{soliton})
exist. No bifurcations of zero eigenvalues of ${\cal L}_0$ occur in
$\mbox{\boldmath $\beta$} \in {\cal B}$.

The kernel of ${\cal L}_1$ has always the eigenvector ${\bf
\Phi}'(x)$. When $z({\cal L}_1) = 1$, this is the only eigenvector
in the kernel of ${\cal L}_1$. When $z({\cal L}_1) > 1$, additional
linearly independent eigenvectors ${\bf u}_0(x)$ exist in the kernel
of ${\cal L}_1$, such that the geometric multiplicity of $\lambda =
0$ in the linearization problem (\ref{eigenvalue-problem}) exceeds
$(N+1)$. Under parameter continuation, the zero eigenvalue $\lambda
= 0$ generally moves either to the real or purely imaginary axes of
$\lambda$. We study this bifurcation in the case, when $z({\cal
L}_1) = 2$, $z({\cal U}) = 0$, and $\mbox{\boldmath $\beta$} \in
{\cal B}$.

\renewcommand{\theLemma}{A.1}
\begin{Proposition}
\label{proposition-zero-eigenvalue} Let $\epsilon$ be the
bifurcation parameter and, at $\epsilon = 0$, there exists a
non-zero eigenvector ${\bf u}_0 \in X_c^{(u)}(\R)$, such that ${\cal
L}_1 {\bf u}_0 = {\bf 0}$ and ${\bf u}_0$ is linearly independent of
${\bf \Phi}'(x)$. Assume that ${\cal L}_1(\epsilon)$ and ${\cal
L}_0(\epsilon)$ are $C^1$-functions at $\epsilon = 0$, such that
$l_0 = \langle {\bf u}_0, {\cal L}_0^{-1}(0) {\bf u}_0 \rangle \neq
0$ and $\delta l_1 = \langle {\bf u}_0, {\cal L}_1'(0) {\bf u}_0
\rangle \neq 0$. Then, there exists $\epsilon_0
> 0$ such that the linearized problem (\ref{L}) has a real
positive eigenvalue $\lambda$ in the domain:
$$
{\cal D}_{\epsilon} = \left\{ \epsilon : 0 < |\epsilon| <
\epsilon_0, \quad {\rm sign}(\epsilon) = - {\rm sign}(l_0 \delta
l_1) \right\}.
$$
\end{Proposition}

\begin{Proof}
We expand solutions of (\ref{diagonalizationproblem}) in power
series of $\epsilon$:
\begin{equation}
\label{power-series-1} {\bf u}(x) = {\bf u}_0(x) + \epsilon {\bf
u}_1(x) + {\rm O}(\epsilon^2), \qquad \gamma = \epsilon \gamma_1 +
{\rm O}(\epsilon^2).
\end{equation}
The function ${\bf u}_1(x)$ solves the non-homogeneous problem in
$X_c^{(u)}({\R})$:
\begin{equation}
\label{bifurcation0} {\cal L}_1(0) {\bf u}_1 + {\cal L}_1'(0) {\bf
u}_0 = \gamma_1 {\cal L}_0^{-1}(0) {\bf u}_0.
\end{equation}
Using the Fredholm Alternative Theorem, we find from
(\ref{bifurcation0}) that $\delta l_1 = \gamma_1 l_0$. If $\delta
l_1 \neq 0$, $l_0 \neq 0$, and ${\rm sign}(\epsilon) = - {\rm
sign}(l_0 \delta l_1)$, the eigenvalue $\gamma$ is negative in the
first order of $\epsilon$, such that $\lambda = \pm \sqrt{-\gamma}$
are real.
\end{Proof}

\begin{Corollary}
\label{corollary-1} Let ${\cal L}_0$ be positive definite, such that
$l_0 > 0$. A new negative eigenvalue $\mu(\epsilon)$ of ${\cal L}_1$
as $\epsilon \neq 0$ results in a new negative eigenvalue
$\gamma(\epsilon)$ of the problem (\ref{diagonalizationproblem}),
such that
\begin{equation}
\label{corollary217} \lim_{\epsilon \to 0} \frac{\gamma}{\mu} =
\frac{\langle {\bf u}_0, {\bf u}_0 \rangle}{ \langle {\bf u}_0,
{\cal L}_0^{-1}(0) {\bf u}_0 \rangle}.
\end{equation}
\end{Corollary}

Bifurcation $z({\cal L}_1) > 1$ may occur only if $N > 1$ in the
system (\ref{NLS}). Analysis of this bifurcation with the
Lyapunov-Schmidt reduction method are reported in a similar content
in \cite{KK04,KKS04,Sand}.

\subsection{Zero eigenvalues of ${\cal U}$}

When $z({\cal U}) > 0$, algebraic multiplicity of $\lambda = 0$
exceeds $(2N + 2)$. Under a parameter continuation, the zero
eigenvalue $\lambda = 0$ generally moves either to the real or
purely imaginary axis of $\lambda$. We study this bifurcation in the
case, when $z({\cal U}) = 1$, $z({\cal L}_1) = 1$, and
$\mbox{\boldmath $\beta$} \in {\cal B}$.

\renewcommand{\theLemma}{A.2}
\begin{Proposition}
\label{proposition-3} Let $\epsilon$ be the bifurcation parameter
and, at $\epsilon = 0$, there exists a non-zero eigenvector
$\mbox{\boldmath $\nu$} \in \R^N$, ${\cal U} \mbox{\boldmath $\nu$}
= {\bf 0}$, such that the eigenvector ${\bf u}_0 \in X_c^{(u)}(\R)$
solves the problem:
\begin{equation}
\label{aux-problem} {\cal L}_1 {\bf u}_0 = - \sum_{n=1}^N \nu_n
\Phi_n(x) {\bf e}_n.
\end{equation}
Assume that ${\cal L}_1(\epsilon)$, ${\cal L}_0(\epsilon)$, and
${\cal U}(\epsilon)$ are $C^1$-functions at $\epsilon = 0$, such
that $l_0 = \langle {\bf u}_0, {\cal L}_0^{-1}(0) {\bf u}_0 \rangle
\neq 0$ and $\delta u = \langle \mbox{\boldmath $\nu$}, {\cal U}'(0)
\mbox{\boldmath $\nu$} \rangle \neq 0$. Then, there exists
$\epsilon_0 > 0$ such that the linearized problem (\ref{L}) has a
real positive eigenvalue $\lambda$ in the domain:
$$
{\cal D}_{\epsilon} = \left\{ \epsilon : 0 < |\epsilon| <
\epsilon_0, \quad {\rm sign}(\epsilon) = - {\rm sign}(l_0 \delta u)
\right\}.
$$
\end{Proposition}

\begin{Proof}
If there exists $\mbox{\boldmath $\nu$} \in \R^N$, such that ${\cal
U}(0) \mbox{\boldmath $\nu$} = {\bf 0}$, then the eigenvector ${\bf
u}_0(x)$ for the problem (\ref{aux-problem}) is given explicitly as
\begin{equation}
\label{bifurcation2} {\bf u}_0(x) = \sum_{n=1}^N \nu_n
\frac{\partial {\bf \Phi}(x)}{\partial \beta_n}.
\end{equation}
Using ${\cal L}_1 \partial {\bf \Phi}/\partial \beta_n = - \Phi_n
{\bf e}_n$ and ${\cal L}_0 \Phi_n {\bf e}_n = {\bf 0}$ for any
$\epsilon$, we obtain the following derivative relations:
\begin{eqnarray}
\label{relation-epsilon-1} {\cal L}_1(0) \frac{\partial
\mbox{\boldmath $\Phi$}'(0)}{\partial \beta_n} + {\cal L}_1'(0)
\frac{\partial \mbox{\boldmath $\Phi$}}{
\partial \beta_n} & = & - \Phi_n'(0) {\bf e}_n, \\
\label{relation-epsilon-2} {\cal L}_0(0) \Phi_n'(0) {\bf e}_n +
{\cal L}_0'(0) \Phi_n {\bf e}_n & = & 0,
\end{eqnarray}
where ${\bf \Phi}'(0)$ stands for derivative of ${\bf
\Phi}(\epsilon)$ in $\epsilon$. We expand solutions of
(\ref{diagonalizationproblem}) in power series of $\epsilon$:
\begin{equation}
\label{power-series-2} {\bf u}(x) = {\bf u}_0(x) + \epsilon {\bf
u}_1(x) + {\rm O}(\epsilon^2), \qquad \gamma = \epsilon \gamma_1 +
{\rm O}(\epsilon^2).
\end{equation}
Since the eigenvector ${\bf u}_0(x)$ solves the non-homogeneous
problem (\ref{aux-problem}), the relation between ${\bf u}(x)$ and
${\bf w}(x)$ is modified as follows:
\begin{equation}
{\bf w} = \lambda {\cal L}_0^{-1} {\bf u} + \frac{1}{\lambda}
\sum_{n=1}^N \nu_n \Phi_n(x) {\bf e}_n.
\end{equation}
As a result, the function ${\bf u}_1(x)$ solves the non-homogeneous
problem:
\begin{equation}
\label{bifurcation3} {\cal L}_1(0) {\bf u}_1 + {\cal L}_1'(0) {\bf
u}_0 = \gamma_1 {\cal L}_0^{-1}(0) {\bf u}_0 - \sum_{n=1}^N \nu_n
\Phi_n'(0) {\bf e}_n,
\end{equation}
subject to the constraints:
\begin{equation}
\label{bifurcation4} \langle \Phi_n(0) {\bf e}_n, {\bf u}_1 \rangle
+ \langle \Phi_n'(0) {\bf e}_n, {\bf u}_0 \rangle = 0.
\end{equation}
Using the Fredholm Alternative Theorem, we find from
(\ref{bifurcation3}) and (\ref{bifurcation4}) that
\begin{equation}
\label{bifurcation100} \langle {\bf u}_0, {\cal L}_1'(0) {\bf u}_0
\rangle = \gamma_1 \langle {\bf u}_0, {\cal L}_0^{-1}(0) {\bf u}_0
\rangle - 2 \sum_{n=1}^N \nu_n \langle \Phi_n'(0) {\bf e}_n, {\bf
u}_0 \rangle.
\end{equation}
As a result,
\begin{equation}
\gamma_1 \langle {\bf u}_0 | {\cal L}_0^{-1}(0) {\bf u}_0 \rangle =
\sum_{n=1}^N \sum_{m=1}^N \nu_{n} \nu_{m} \frac{\partial}{\partial
\epsilon} \langle \Phi_m(\epsilon) {\bf e}_m, \frac{\partial {\bf
\Phi}(\epsilon)}{\partial \beta_n} \rangle \biggr|_{\epsilon = 0} =
\frac{1}{2} \langle \mbox{\boldmath $\nu$}, {\cal U}'(0)
\mbox{\boldmath $\nu$} \rangle,
\end{equation}
such that $\delta u  = 2 \gamma_1 l_0$. When $\delta u \neq 0$, $l_0
\neq 0$, and ${\rm sign}(\epsilon) = - {\rm sign}(l_0 \delta u)$,
the eigenvalue $\gamma$ is negative in the first order of
$\epsilon$, such that $\lambda = \pm \sqrt{-\gamma}$ are real.
\end{Proof}

\begin{Corollary}
\label{corollary-3} Let ${\cal L}_0$ be positive definite, such that
$l_0 > 0$. A new negative eigenvalue $\mu(\epsilon)$ of ${\cal U}$
as $\epsilon \neq 0$ results in a new negative eigenvalue
$\gamma(\epsilon)$ of the problem (\ref{diagonalizationproblem}).
\end{Corollary}

Bifurcation $z({\cal U}) > 0$ was analyzed in \cite{PelKiv,Skr} with
power series expansions of the problem (\ref{L}) near $\lambda = 0$
and in \cite{Br1,KS1} with Taylor series expansions of the Evans
function.

\subsection{Multiple non-zero eigenvalues of zero energy}

When the problem (\ref{L}) has a multiple eigenvalue $\lambda =
\lambda_0 \in i \mathbb{R}$ of zero energy, the corresponding
eigenvector $({\bf u}_0,{\bf w}_0)^T$ satisfies the conditions of
the Fredholm Alternative Theorem: $\langle {\bf u}_0, {\cal L}_1
{\bf u}_0 \rangle = 0$, $\langle {\bf w}_0, {\cal L}_0 {\bf w}_0
\rangle = 0$, such that $h[{\bf u}_0,{\bf w}_0] = 0$ and $l_0 =
\langle {\bf u}_0, {\cal L}_0^{-1} {\bf u}_0 \rangle = 0$. Under
parameter continuations, multiple eigenvalues are generally
destroyed and new complex eigenvalues $\lambda$ may arise in the
problem (\ref{L}). We study this bifurcation in the case, when a
multiple eigenvalue $\lambda = \lambda_0$ has algebraic multiplicity
two and geometric multiplicity one, while $\mbox{\boldmath $\beta$}
\in {\cal B} \cup \partial {\cal B}$.

\renewcommand{\theLemma}{A.3}
\begin{Proposition}
\label{proposition-5} Let $\epsilon$ be the bifurcation parameter
and, at $\epsilon = 0$, there exist non-zero eigenvectors ${\bf
u}_0, {\bf u}_0' \in X_c^{(u)}(\R)$ for $\gamma_0 \in \R$ and
$\gamma_0 < \beta_{\rm min}^2$, such that
\begin{eqnarray}
{\cal L}_1 {\bf u}_0 & = & \gamma_0 {\cal L}_0^{-1} {\bf u}_0, \\
{\cal L}_1 {\bf u}_0' & = & \gamma_0 {\cal L}_0^{-1} {\bf u}_0' +
{\cal L}_0^{-1} {\bf u}_0,
\end{eqnarray}
and $l_0 = \langle {\bf u}_0, {\cal L}_0^{-1} {\bf u}_0 \rangle =
0$. Assume that ${\cal L}_1(\epsilon)$ and ${\cal L}_0(\epsilon)$
are $C^1$-functions at $\epsilon = 0$, such that $l_0' = \langle
{\bf u}_0, {\cal L}_0^{-1}(0) {\bf u}_0' \rangle \neq 0$ and $\delta
h = \langle {\bf u}_0, \left( {\cal L}_1'(0) - \gamma_0 {\cal
L}_0^{-1 \prime}(0) \right) {\bf u}_0 \rangle \neq 0$. Then, there
exists $\epsilon_0 > 0$ such that the linearized problem (\ref{L})
has two complex eigenvalues $\lambda$ with ${\rm Re}(\lambda) > 0$
in the domain:
$$
{\cal D}_{\epsilon} = \left\{ \epsilon : 0 < |\epsilon| <
\epsilon_0, \quad {\rm sign}(\epsilon) = - {\rm sign}(l_0' \delta h)
\right\}.
$$
\end{Proposition}

\begin{Proof}
We expand solutions of (\ref{diagonalizationproblem}) in power
series of $\epsilon^{1/2}$:
\begin{eqnarray}
{\bf u}(x) & = & {\bf u}_0(x) + \epsilon^{1/2} \gamma_1 {\bf
u}_0'(x) + \epsilon {\bf u}_2(x) + {\rm O}(\epsilon^{3/2}), \\
\gamma & = & \gamma_0 + \epsilon^{1/2} \gamma_1 + \epsilon \gamma_2
+ {\rm O}(\epsilon^{3/2}).
\end{eqnarray}
The function ${\bf u}_2(x)$ solves the non-homogeneous problem in
$X_c^{(u)}(\R)$:
\begin{eqnarray}
\label{bifurcation13} {\cal L}_1(0) {\bf u}_2 + {\cal L}_1'(0) {\bf
u}_0 = \gamma_0 {\cal L}_0^{-1}(0) {\bf u}_2 + \gamma_0 {\cal
L}_0^{-1\prime}(0) {\bf u}_0 + \gamma_1^2 {\cal L}_0^{-1}(0) {\bf
u}_0' + \gamma_2 {\cal L}_0^{-1}(0) {\bf u}_0.
\end{eqnarray}
Using the Fredholm Alternative Theorem, we find from
(\ref{bifurcation13}) that $\delta h = \gamma_1^2 l_0'$. Since
$(\gamma - \gamma_0)^2 = \epsilon \gamma_1^2 + {\rm
O}(\epsilon^{3/2})$, the eigenvalues $\gamma$ bifurcate into the
complex plane if $l_0' \neq 0$, $\delta h \neq 0$, and ${\rm
sign}(\epsilon) = - {\rm sign}(l_0' \delta h)$.
\end{Proof}

\begin{Corollary}
\label{corollary-5} Let $l_0' \neq 0$ and $\delta h \neq 0$. There
exists $\epsilon_0 > 0$ such that the problem
(\ref{diagonalizationproblem}) has two real eigenvalues $\gamma$ in
the domain,
$$
{\cal D}_{\epsilon} = \left\{ \epsilon : 0 < |\epsilon| <
\epsilon_0, \quad {\rm sign}(\epsilon) = {\rm sign}(l_0' \delta h)
\right\}
$$
with oppositely signed quadratic forms
$$
\langle {\bf u}, {\cal L}_0^{-1} {\bf u} \rangle = 2 \epsilon^{1/2}
\gamma_1 l_0' + {\rm O}(\epsilon).
$$
\end{Corollary}

When $0 < \gamma_0 < \beta_{\rm min}$, multiple eigenvalue $\lambda
= \lambda_0$ is purely imaginary, and the bifurcation of Proposition
\ref{proposition-5} is the instability bifurcation. When $\gamma_0 <
0$, the eigenvalue $\lambda = \lambda_0$ is purely real and is thus
already unstable. Characteristic features of bifurcation of multiple
eigenvalues were analyzed in \cite{G2}. This bifurcation is generic
when purely imaginary eigenvalues of positive and negative energies
$h[{\bf u},{\bf w}]$ coalesce, according to Corollary
\ref{corollary-5} \cite{Skr10}. General results on the collisions of
purely imaginary eigenvalues of different energies and the concept
of so-called Krein signatures can be found in \cite{MacKay}.

\subsection{Embedded eigenvalues}

When the problem (\ref{L}) has an embedded eigenvalue $\lambda =
\lambda_0 \in i \mathbb{R}$ with $|{\rm Im}(\lambda_0)| > \beta_{\rm
min}$, it is generically unstable under parameter continuation
\cite{G2,LP1,LP2}. When it has a positive energy, it disappears from
the continuous spectrum, while when it has a negative energy, it
bifurcates as complex unstable eigenvalues with ${\Re}(\lambda) > 0$
\cite{CPV04}. We study this bifurcation in the case, when an
embedded eigenvalue has the geometric and algebraic multiplicities
one, while $\mbox{\boldmath $\beta$} \in {\cal B} \cup \partial
{\cal B}$.

\renewcommand{\theLemma}{A.4}
\begin{Proposition}
\label{proposition-7} Let $\epsilon$ be the bifurcation parameter
and, at $\epsilon = 0$, there exist a non-zero eigenvector ${\bf
u}_0 \in X_c^{(u)}(\R)$ for $\gamma_0 \in \R$, $\gamma_0 >
\beta_{\rm min}^2$, such that
\begin{eqnarray}
{\cal L}_1 {\bf u}_0 = \gamma_0 {\cal L}_0^{-1} {\bf u}_0.
\end{eqnarray}
Assume that ${\cal L}_1(\epsilon)$ and ${\cal L}_0(\epsilon)$ are
$C^1$-functions at $\epsilon = 0$, such that $l_0 = \langle {\bf
u}_0, {\cal L}_0^{-1}(0) {\bf u}_0 \rangle < 0$ and $\Gamma \neq 0$
in (\ref{Fermat}) below. Then, there exists $\epsilon_0
> 0$ such that the linearized problem (\ref{L}) has two complex
eigenvalues $\lambda$ with ${\rm Re}(\lambda) > 0$.
\end{Proposition}

\begin{Proof}
For embedded eigenvalues, we use the linearized problem in the
original form (\ref{L}). Consider perturbation series expansions
near the embedded eigenvalue $\lambda = \lambda_0$, with ${\rm
Im}(\lambda_0) > \beta_{\rm min}$:
\begin{eqnarray}
\label{pertur101} {\bf u}(x) & = & {\bf u}_0(x) + \epsilon {\bf
u}_1(x) + \epsilon^2 {\bf u}_2(x) + {\rm O}(\epsilon^3), \\
\label{pertur102} {\bf w}(x) & = & {\bf w}_0(x) + \epsilon {\bf
w}_1(x) + \epsilon^2 {\bf w}_2(x) + {\rm O}(\epsilon^3),
\end{eqnarray}
and
\begin{equation}
\label{pertur103} \lambda = \lambda_0 + \epsilon \lambda_1 +
\epsilon^2 \lambda_2 + {\rm O}(\epsilon^3).
\end{equation}
Corrections of the perturbation series (\ref{pertur101}) and
(\ref{pertur102}) satisfy linear non-homogeneous equations following
from the linearized problem (\ref{L}):
\begin{eqnarray}
\nonumber {\cal L}_1(0) {\bf u}_1 + \lambda_0 {\bf w}_1
& = & - {\cal L}_1'(0) {\bf u}_0 - \lambda_1 {\bf w}_0, \\
\label{non104} {\cal L}_0(0) {\bf w}_1 - \lambda_0 {\bf u}_1 & = & -
{\cal L}_0'(0) {\bf w}_0 + \lambda_1 {\bf u}_0
\end{eqnarray}
and
\begin{eqnarray}
\nonumber {\cal L}_1(0) {\bf u}_2 + \lambda_0 {\bf w}_2 & = & -
{\cal L}_1'(0) {\bf u}_1 - \frac{1}{2} {\cal L}_1''(0) {\bf u}_0
- \lambda_1 {\bf w}_1 - \lambda_2 {\bf w}_0, \\
\label{non105} {\cal L}_0(0) {\bf w}_2 - \lambda_0 {\bf u}_2 & = & -
{\cal L}_0'(0) {\bf w}_1 - \frac{1}{2} {\cal L}_0''(0) {\bf w}_0 +
\lambda_1 {\bf u}_1 + \lambda_2 {\bf u}_0.
\end{eqnarray}
Bounded solutions of the problem (\ref{non104}) exist only if the
right-hand-side is orthogonal to the eigenvector $({\bf u}_0,{\bf
w}_0)$. The solvability condition results in the equation:
\begin{eqnarray}
\nonumber \lambda_1 \left( \langle {\bf w}_0, {\bf u}_0 \rangle -
\langle {\bf u}_0, {\bf w}_0 \rangle \right) = \langle {\bf u}_0,
{\cal L}_1'(0) {\bf u}_0 \rangle + \langle {\bf w}_0, {\cal L}_0'(0)
{\bf w}_0 \rangle,
\end{eqnarray}
such that ${\rm Re}(\lambda_1) = 0$. Since the eigenvalue $\lambda =
\lambda_0$ belongs to the continuous spectrum of the problem
(\ref{L}), the correction terms $({\bf u}_1,{\bf w}_1)$ have
non-vanishing tails in the limit $|x| \to \infty$. Assuming that
$\beta_1 \leq \beta_2 \leq ... \leq \beta_N$, we add Sommerfeld
radiation conditions to uniquely determine the correction terms
$({\bf u}_1,{\bf w}_1)$:
\begin{equation}
\label{non106} \left( \begin{array}{c} {\bf u}_1 \\
{\bf w}_1 \end{array} \right) \rightarrow \sum_{j=1}^{K_{\lambda_0}}
g_j^{\pm} \left(
\begin{array}{cc} {\bf e}_j \\ i {\bf e}_j \end{array} \right)
e^{\mp i k_j x}, \qquad x \to \pm \infty,
\end{equation}
where $g_j^{\pm}$ are some constants, $k_j = \sqrt{({\rm
Im}(\lambda_0) - \beta_j)/d_j}$, and $K_{\lambda_0}$ is the number
of branches with $k_j \in \R$. It follows from (\ref{non104}) that
\begin{eqnarray}
\nonumber & \phantom{t} & {\rm Im}\left( \langle {\bf u}_1, {\cal
L}_1'(0) {\bf u}_0 \rangle + \langle {\bf w}_1, {\cal L}_0'(0) {\bf
w}_0 \rangle + \lambda_1 \langle {\bf u}_1, {\bf w}_0 \rangle
- \lambda_1 \langle {\bf w}_1, {\bf u}_0 \rangle \right) \\
\nonumber & = & -\frac{1}{2i} \left( \langle {\bf u}_1, {\cal L}_1
{\bf u}_1 \rangle + \langle {\bf w}_1, {\cal L}_0 {\bf w}_1 \rangle
- \langle {\cal L}_1 {\bf u}_1, {\bf u}_1 \rangle -
\langle {\cal L}_0 {\bf w}_1, {\bf w}_1 \rangle \right) \\
\label{Fermat} & = & -2 \sum_{j=1}^{K_{\lambda_0}} d_j k_j \left(
|g^+_j|^2 + |g^-_j|^2 \right) \equiv \Gamma \leq 0.
\end{eqnarray}
Again, bounded solutions of the problem (\ref{non105}) exist only if
\begin{eqnarray}
\nonumber \lambda_2 \left( \langle {\bf w}_0, {\bf u}_0 \rangle -
\langle {\bf u}_0, {\bf w}_0 \rangle \right) + \lambda_1 \left(
\langle {\bf w}_0, {\bf u}_1 \rangle -
\langle {\bf u}_0, {\bf w}_1 \rangle \right) \\
= \langle {\bf u}_0, {\cal L}_1'(0) {\bf u}_1 \rangle + \langle {\bf
w}_0, {\cal L}_0'(0) {\bf w}_1 \rangle + \frac{1}{2} \langle {\bf
u}_0, {\cal L}_1''(0) {\bf u}_0 \rangle + \frac{1}{2} \langle {\bf
w}_0, {\cal L}_0''(0) {\bf w}_0 \rangle,
\end{eqnarray}
such that
\begin{equation}
\label{pertur111} {\rm Re} (\lambda_2) = \frac{{\rm Im} \left(
\langle {\bf u}_1, {\cal L}_1'(0) {\bf u}_0 \rangle + \langle {\bf
w}_1, {\cal L}_0'(0) {\bf w}_0 \rangle \right)}{ 2 {\rm
Im}(\lambda_0) \langle {\bf u}_0, {\cal L}_0^{-1} {\bf u}_0 \rangle}
= - \sum_{j=1}^{K_{\lambda_0}} \frac{d_j k_j \left( |g_j^+|^2 +
|g_j^-|^2 \right)}{ {\rm Im}(\lambda_0) \langle {\bf u}_0, {\cal
L}_0^{-1} {\bf u}_0 \rangle}.
\end{equation}
When $\Gamma \neq 0$ and $l_0 = \langle {\bf u}_0, {\cal
L}_0^{-1}(0) {\bf u}_0 \rangle < 0$, the embedded eigenvalue
$\lambda = \lambda_0$ becomes a complex unstable eigenvalue
$\lambda$ with ${\rm Re}(\lambda) > 0$.
\end{Proof}

\begin{Corollary}
The linearized problem (\ref{L}) does not have complex or embedded
eigenvalues $\lambda$ if $l_0 = \langle {\bf u}_0, {\cal
L}_0^{-1}(0) {\bf u}_0 \rangle > 0$ and $\Gamma \neq 0$.
\end{Corollary}

\begin{Proof}
The formal computation in (\ref{pertur111}) predicts that ${\rm
Re}(\lambda_2) < 0$, when $\Gamma \neq 0$ and $l_0 > 0$. However,
the correction terms $({\bf u}_1,{\bf w}_1)^T$ in (\ref{non106})
grow exponentially in $x$, since $k_j(\lambda) = \sqrt{({\rm
Im}(\lambda) - i {\rm Re}(\lambda) - \beta_j)/d_j}$ implies that
${\rm Im}(k_j)
> 0$. The embedded eigenvalue $\lambda = \lambda_0$ becomes a
resonant pole with ${\rm Re}(\lambda) < 0$.
\end{Proof}

Characteristic features of bifurcations of embedded eigenvalues were
analyzed in \cite{CPV04,G2}. This bifurcation is generic for
multi-hump vector solitons at the boundaries of the existence domain
$\mbox{\boldmath $\beta$} \in \partial {\cal B}$ \cite{TS02}.

\vspace{1cm}

Summarizing, there exist four bifurcations, which may lead to
unstable eigenvalues in the spectral problem (\ref{L}): (i) $z({\cal
L}_1) > 1$, (ii) $z({\cal U}) > 0$, (iii) multiple eigenvalue
$\lambda_0 \in i \R$, $|{\rm Im}(\lambda_0)| < \beta_{\rm min}$, and
(iv) embedded eigenvalue $\lambda_0 \in i\R$, $|{\rm Im}(\lambda_0)|
> \beta_{\rm min}$. Let $n_X(h)$ be the negative index of the
linearized Hamiltonian in the constrained space $X_c^{(u)}(\R)$,
i.e. the number of negative eigenvalues of ${\cal L}_1$ and ${\cal
L}_0$ in $X_c^{(u)}(\R)$. It is known from \cite{GSS90} (see also
\cite{P03,PelKiv}) that
\begin{equation}
\label{total-index0} n_X(h) = n({\cal L}_1) - p({\cal U}) + n({\cal
L}_0).
\end{equation}
The closure relation (\ref{closure}) gives then:
\begin{equation}
\label{total-index} n_X(h) = N_{\rm real} + 2 N_{\rm imag}^- + 2
N_{\rm comp}.
\end{equation}
It is clear from (\ref{total-index}) that bifurcations (i) and (ii)
change the negative index $n_X(h)$ due to a change in $N_{\rm real}$
by one, while bifurcations (iii) and (iv) do not change the negative
index $n_X(h)$ due to an exchange in $N_{\rm imag}^-$ and $N_{\rm
comp}$.

\newpage
\begin{figure}[htbp]
\begin{center}
\includegraphics[height=1.46in]{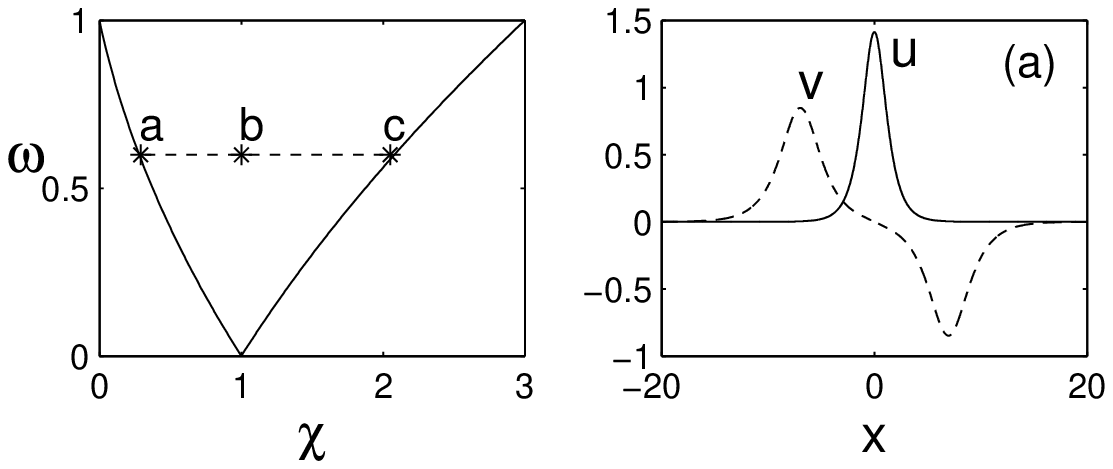}

\vspace{0.4cm}
\includegraphics[height=1.33in]{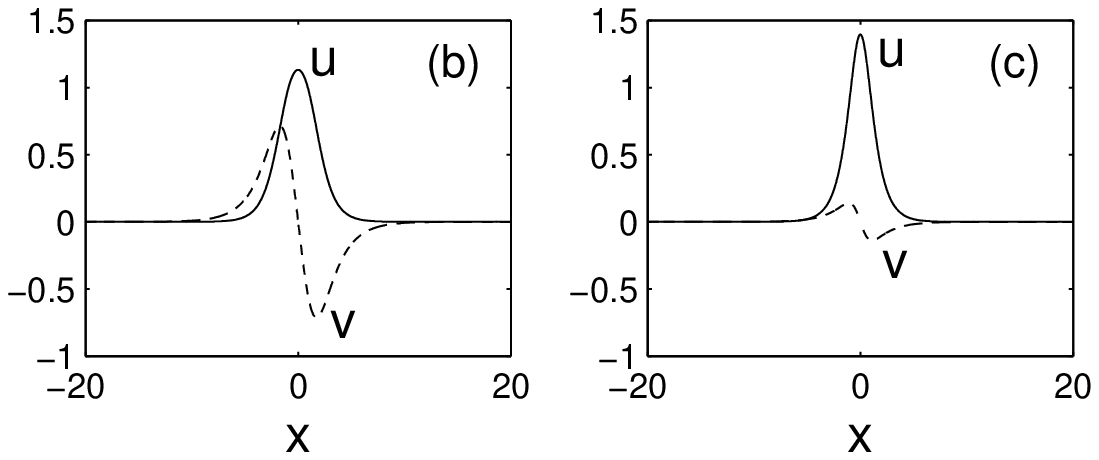}
\end{center}
\caption{Vector solitons of the 1st-family in the cubic model
(\ref{CPNLS}) at $\beta=0.36$ and three $\chi$ values marked as "*"
in the upper left figure. Here $\omega\equiv \sqrt{\beta}$, and $(u,
v)\equiv (\Phi_1, \Phi_2)$. }
\end{figure}

\begin{figure}[htbp]
\begin{center}
\includegraphics[height=2.4in]{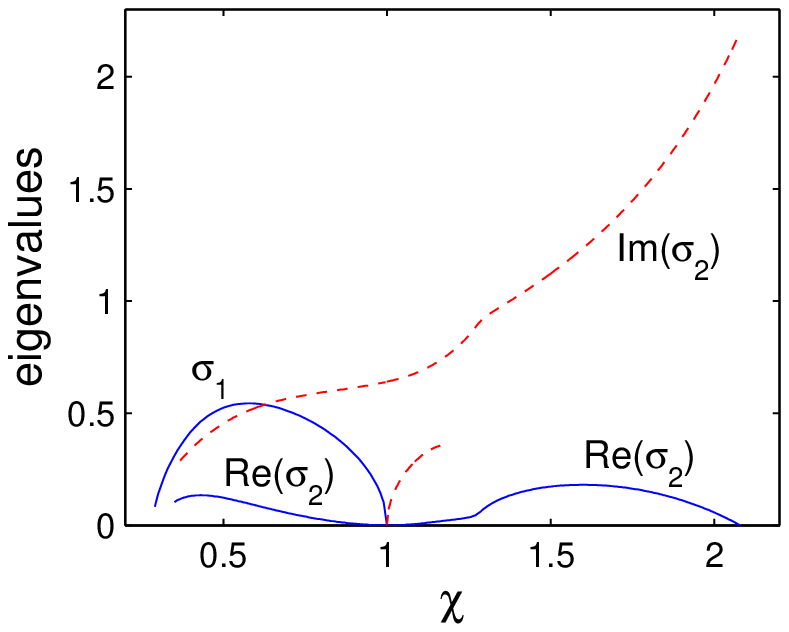}
\end{center}
\caption{Eigenvalue spectrum of the 1st family of vector solitons in
the cubic model (\ref{CPNLS}) at $\beta=0.36$. The solid (dashed)
lines are the real (imaginary) parts of the eigenvalues.  }
\end{figure}

\begin{figure}[htbp]
\begin{center}
\includegraphics[height=1.5in]{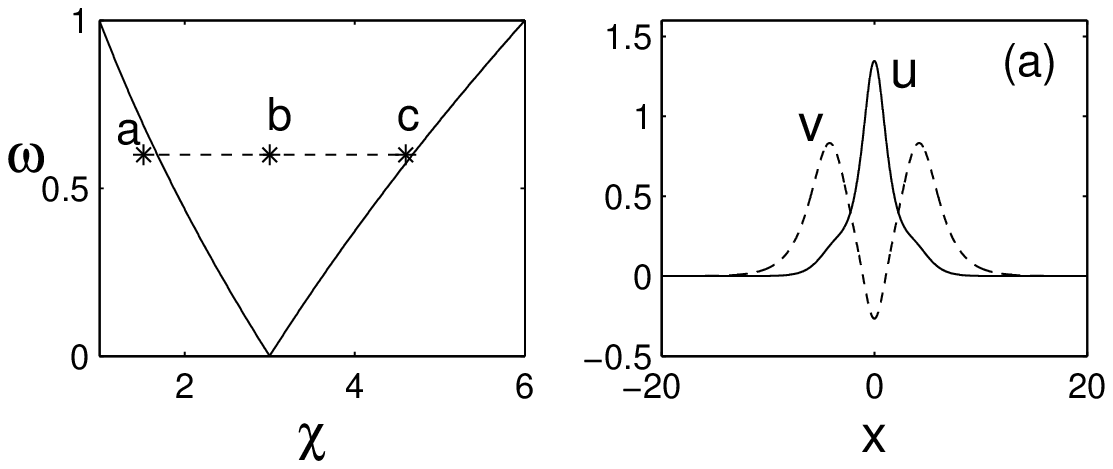}

\vspace{0.4cm}
\includegraphics[height=1.31in]{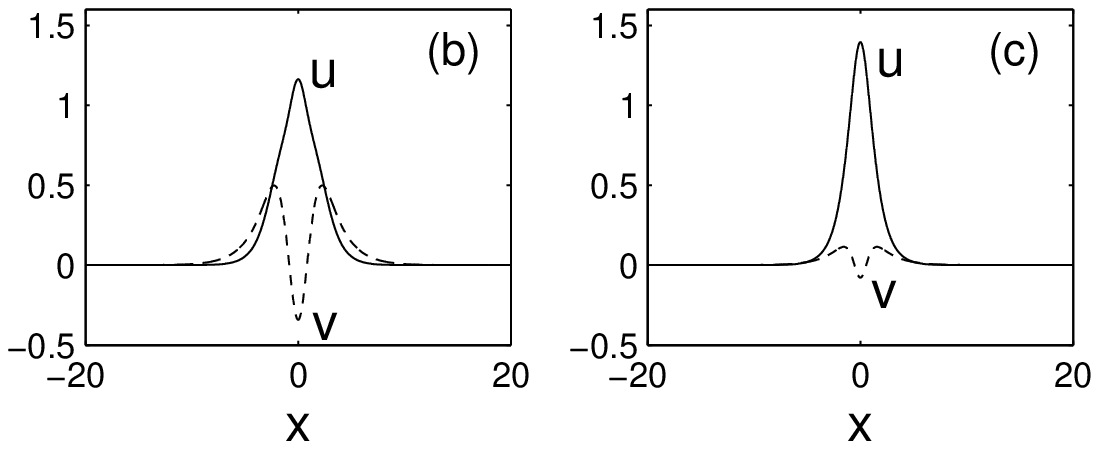}
\end{center}
\caption{Vector solitons of the 2nd family in the cubic model
(\ref{CPNLS}) at $\beta=0.36$ and three $\chi$ values. Notations
$\omega, u$ and $v$ are the same as in Fig. 1.}
\end{figure}

\begin{figure}[htbp]
\begin{center}
\includegraphics[height=2.4in]{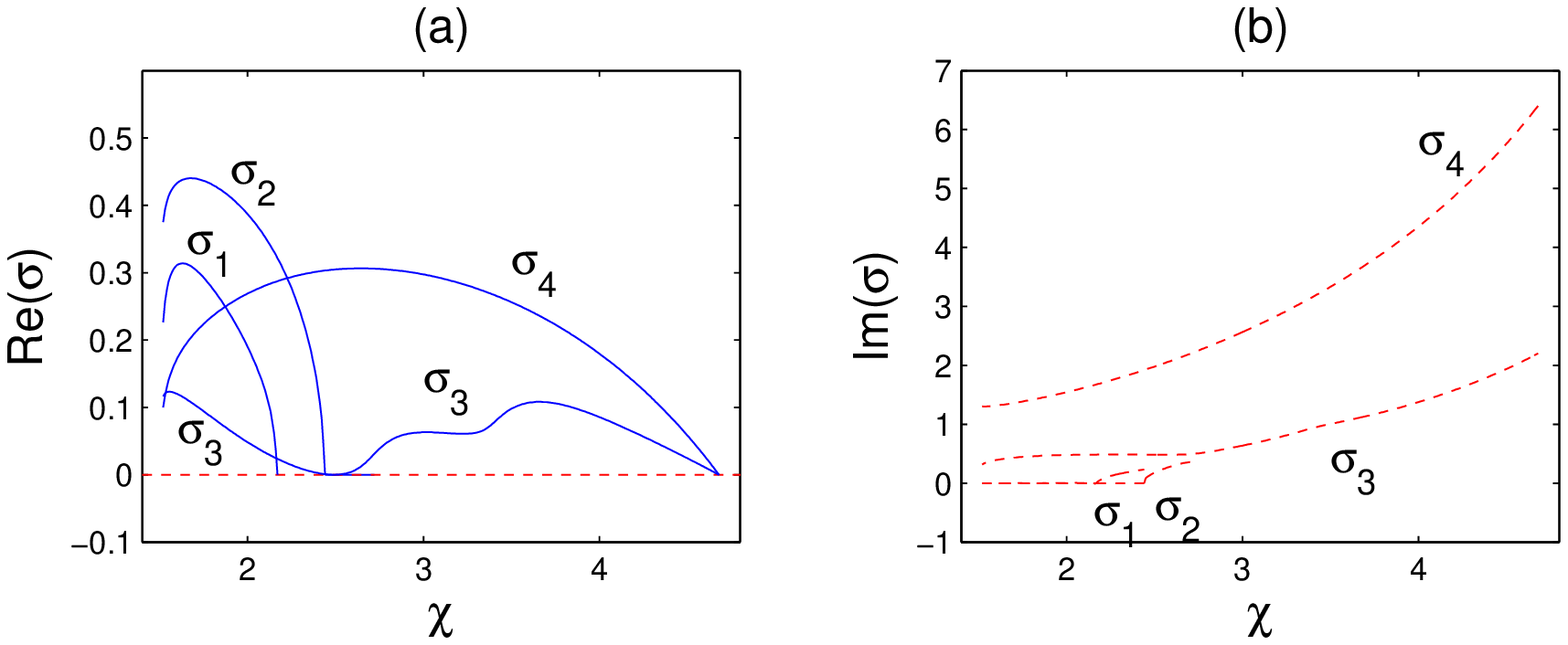}
\end{center}
\caption{Eigenvalue spectrum of the 2nd family of vector solitons in
the cubic model (\ref{CPNLS}) at $\beta=0.36$. The solid (dashed)
lines are the real (imaginary) parts of the eigenvalues.}
\end{figure}

\begin{figure}[htbp]
\begin{center}
\includegraphics[height=2.5in]{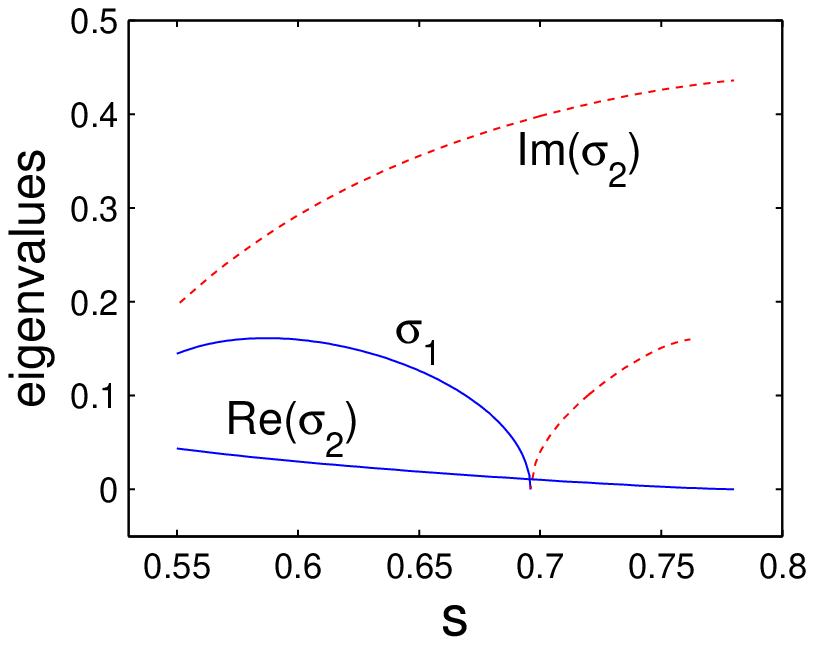}
\end{center}
\caption{Eigenvalue spectrum of the 2nd family of
vector solitons in the saturable model (\ref{Saturable}) at $\beta=0.16$.
The solid (dashed) lines are the real (imaginary) parts of the eigenvalues. }
\end{figure}

\begin{figure}[htbp]
\begin{center}
\includegraphics[height=2.44in]{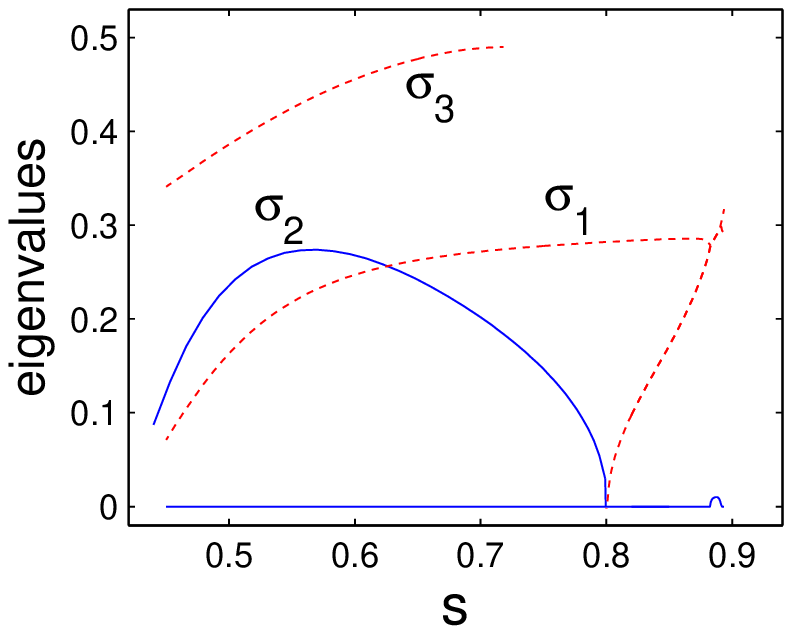}
\includegraphics[height=2.5in]{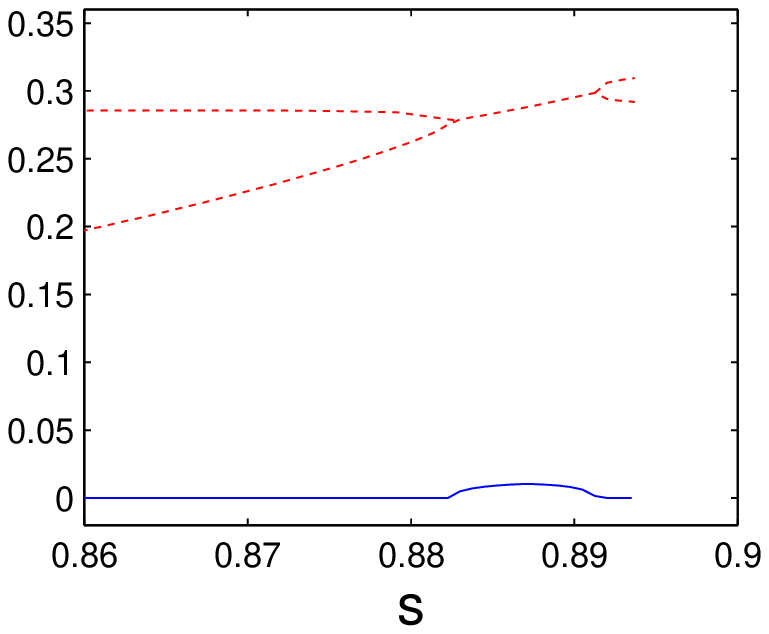}
\end{center}
\caption{Eigenvalue spectrum of the 2nd family of vector solitons in
the saturable model (\ref{Saturable}) at $\beta=0.49$. The solid
(dashed) lines are the real (imaginary) parts of the eigenvalues.
The right figure is a zoom-in of the left figure.}
\end{figure}

\begin{figure}[htbp]
\begin{center}
\includegraphics[height=2.58in]{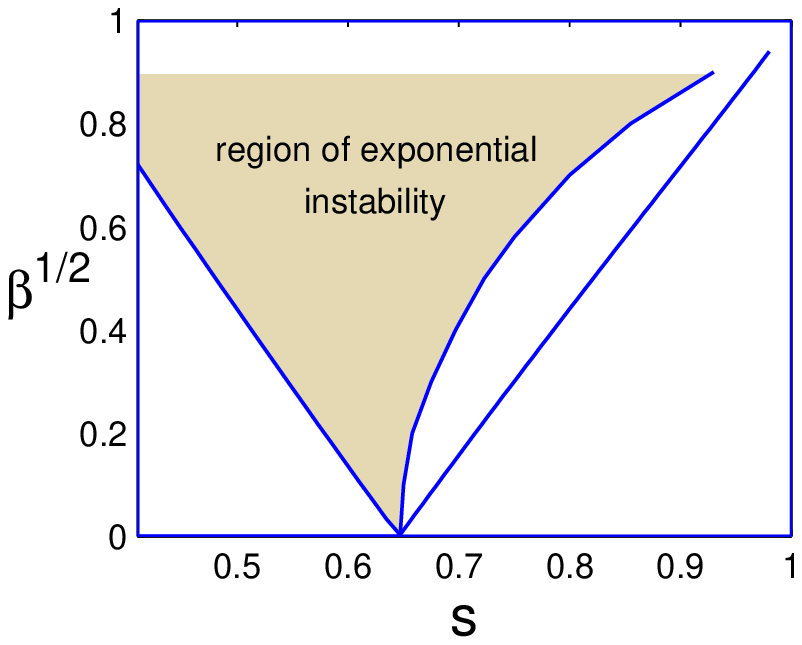}
\includegraphics[height=2.5in]{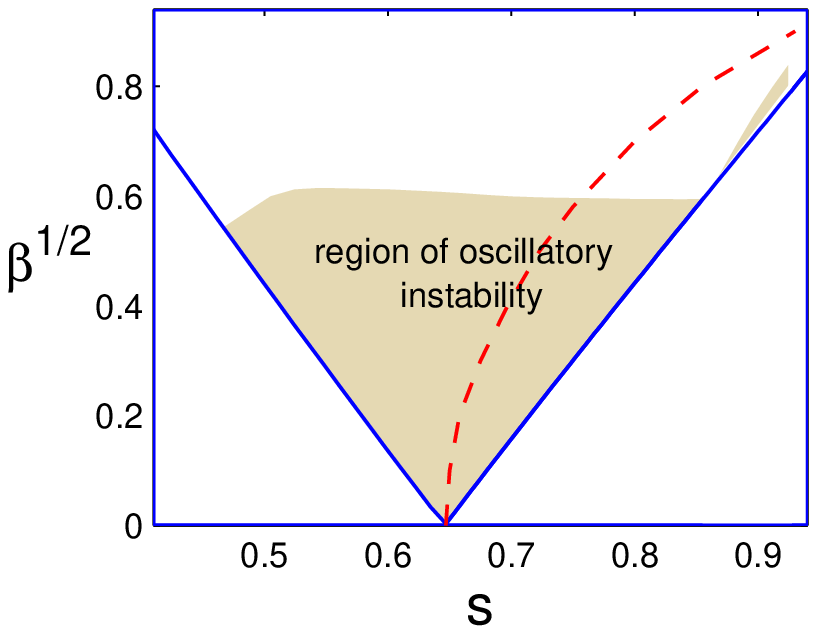}
\end{center}
\caption{Regions (shaded) of exponential and oscillatory
instabilities of the 2nd family of vector solitons in the
saturable model (\ref{Saturable}).}
\end{figure}

\end{document}